# Characterization of OCS-HCCCCH and N$_2$O-HCCCCH dimers: Theory and experiment

A.J. Barclay,[a] A. Pietropolli Charmet,[b] K.H. Michaelian,[c] and N. Moazzen-Ahmadi[a]

[a] *Department of Physics and Astronomy, University of Calgary, 2500 University Drive North West, Calgary, Alberta T2N 1N4, Canada*

[b] *Dipartimento di Scienze Molecolari e Nanosistemi, Università Ca' Foscari Venezia, Via Torino 155, I-30172, Mestre, Venezia, Italy*

[c] *CanmetENERGY, Natural Resources Canada, One Oil Patch Drive, Devon, Alberta T9G 1A8, Canada*

Address for correspondence: Prof. N. Moazzen-Ahmadi,
               Department of Physics and Astronomy,
               University of Calgary,
               2500 University Drive North West,
               Calgary, Alberta T2N 1N4 Canada.

**Abstract**

The infrared spectra of the weakly-bound dimers OCS-HCCCCH, in the region of the $\nu_1$ fundamental band of OCS (2050 cm$^{-1}$), and N$_2$O-HCCCCH, in the region of the $\nu_1$ fundamental band of N$_2$O (2200 cm$^{-1}$), were observed in a pulsed supersonic slit jet expansion probed with tunable diode/QCL lasers. Both OCS-HCCCCH and N$_2$O-HCCCCH were found to have planar structure with side-by-side monomer units having nearly parallel axes. These bands have hybrid rotational structure which allows for estimates of the orientation of OCS and N$_2$O in the plane of their respective dimers. Analogous bands for OCS-DCCCCD and N$_2$O-DCCCCD were also observed and found to be consistent with the normal isotopologues. Various levels of *ab initio* calculations were performed to find stationary points on the potential energy surface, optimized structures and interaction energies. Four stable geometries were found for OCS-HCCCCH and three for N$_2$O-HCCCCH. The rotational parameters at CCSD(T*)-F12c level of theory give results in very good agreement with those obtained from the observed spectra. In both dimers, the experimental structure corresponds to the lowest energy isomer.

**Introduction**

Acetylene is a prototype molecule for examining weak intermolecular forces involving carbon-carbon triple bond. Pure and mixed acetylene clusters have been studied extensively both theoretically and experimentally [1, 2, 3, 4]. In particular, mixed dimers containing OCS and $N_2O$ with acetylene are well characterized using microwave and infrared spectroscopy. To date, only one isomer of $N_2O$-HCCH has been observed in the gas phase [3, 5, 6]. This isomer, which is almost certainly the most stable structure, is planar with near parallel monomer units. For OCS-HCCH two isomers have been detected; a planar near-parallel structure analogous to the observed $N_2O$-HCCH and a T-shaped isomer having $C_{2v}$ symmetry, with OCS forming the stem of the T, and the S atom in the inner position pointing to the triple bond of $C_2H_2$ [3, 7, 8].

The longer counterpart of acetylene, diacetylene, provides a simple system with multiple triple-bond sites; thus clusters containing diacetylene potentially display much richer-structured energy surfaces. For example, theoretical calculations in Ref. [9] identify three structural isomers, a Y-shaped π-type hydrogen-bonded structure with $C_S$ symmetry, a parallel slipped structure with $C_{2h}$ symmetry, and a cross shaped structure with $D_{2d}$ symmetry. Since this dimer has not been observed experimentally, nor has its complete intermolecular potential surface been investigated with high-level *ab initio* calculations, it is unclear what type of tunneling dynamics exist in $(HCCCCH)_2$ and if there are out of plane paths via the cross-shaped structure, or the situation is similar to acetylene dimer where the large amplitude tunneling takes place in the plane of the dimer.

There are relatively few gas phase studies of mixed dimers containing diacetylene. The first of these is a microwave study of $NH_3$-HCCCCH and $H_2O$-HCCCCH by Matsumura et al. [10]. Here, $NH_3$-HCCCCH was found to be an axially symmetric complex where a diacetylene hydrogen bonds to the nitrogen in ammonia to form a symmetric top and $H_2O$-HCCCCH has a

$C_{2v}$ planar symmetry where a diacetylene hydrogen bonds to the oxygen of water. The structural similarities with $NH_3$-HCCH and $H_2O$-HCCH suggest that diacetylene closely parallels acetylene in complexes with other bases. On the other hand a dissimilar behavior might be expected when acetylene and diacetylene act as bases in complexes with acids.

A study by Yang et al. [11] of HCN-HCCCCH in the infrared region found only a linear configuration despite the fact that both linear and T-shaped structures are observed in HCN-HCCH and the T-shaped isomer is strongly favored in the molecular beam expansion of HCN-HCCH. Although, this observation seems to confirm the dissimilar behavior between acetylene and diacetylene when they act as bases, Yang et al. [11] attributed the absence of the T-shaped isomer in their experiment to the very asymmetric top character of the HCN-HCCCCH which disperses its spectra and gives rise to an unfavorable partition function and consequently low signal to noise ratio.

In the present paper, we observe and analyze the spectra of OCS-HCCCCH and OCS-DCCCCD, in the region of the $\nu_1$ fundamental of OCS, and $N_2O$-HCCCCH and $N_2O$-DCCCCD, in the region of the $\nu_1$ fundamental of $N_2O$. Both dimers were found to have planar $C_S$ symmetry with monomer axes nearly parallel to each other. Furthermore, the spectra have hybrid a-, b-type rotational structure which allow for good estimates of the orientation of OCS and $N_2O$ in their dimers. The four independent structural constants obtained from the analysis of spectra for OCS-HCCCCH and OCS-DCCCCD are insufficient for accurate determination of an experimental structure. The same can be said in the case of $N_2O$-HCCCCH. Therefore, we carried out various levels of *ab initio* calculations in support of our experimental findings. Counterpoise-corrected binding energy calculations indicate that the observed structures are the most stable isomers. In addition, three less energetically favorable structures were found for OCS-HCCCCH and two for

N$_2$O-HCCCCH. Experimental rotational constants are in excellent agreement with the rotational constants at the CCSD(T*)-F12c level of theory. Our results closely parallel mixed acetylene complexes with OCS and N$_2$O.

1. **Computational details**

To support our experimental analysis, as well as to characterize the stationary points on the potential energy surface (PES) of these complexes, we performed several *ab initio* calculations.

As a first step, we searched the potential energy surface (PES) of both OCS-HCCCCH and N$_2$O-HCCCCH by means of density functional theory (DFT). Two different functionals, namely B3LYP [12, 13] and B2PLYP [14], were used employing the DFT-D3 dispersion corrections proposed by Grimme [15], in conjunction with the maug-cc-pVTZ basis set [16]. The inclusion of these corrections is mandatory since, even if DFT methods are nowadays widely employed for treating a wide variety of chemical problems, ranging from spectroscopic analysis to molecular complexes and adsorption processes (see for example Refs. [17, 18, 19, 20, 21, 22, 23, 24, 25]), the role of dispersion contributions is fundamental for modeling correctly the energetics of van der Waals adducts [26, 27, 28]. The structures of the stationary points for both OCS-HCCCCH and N$_2$O-HCCCCH were first optimized at B3LYP-D3 level of theory and then refined using the B2PLYP-D3 functional. Calculations carried out with both functionals led to identification of four stationary points for OCS-HCCCCH and three for N$_2$O-HCCCCH; subsequent hessian calculations carried out on these optimized structures, corrected by the basis set superposition error (BSSE) using the counterpoise correction (CP) as proposed by Boys and Bernardi [29], confirmed that they correspond to true minima on the PES. These isomers are shown in Fig. 1 for OCS-HCCCCH and in Fig. 2 for N$_2$O-HCCCCH. We computed the corresponding counterpoise corrected (CP) values of both binding and interaction energies (hereafter labeled as BE and IE, respectively) at the B2PLYP-D3 level of theory for all of the structures. We corrected these BE values by also taking into account the zero-point vibrational (ZPV) correction for each species, defined as

$$\Delta E = \frac{1}{2} \sum_n \omega_n \qquad (1)$$

where the harmonic frequencies $\omega_n$ (the index $n$ labels the $n$-th vibrational mode) were obtained at B2PLYP-D3/maug-cc-pVTZ level. To determine a better estimate of the binding energy for all the structures, we performed additional single-point calculations on each of these optimized geometries at the coupled-cluster level of theory using the singles and doubles approximation augmented by a perturbative treatment of triple excitations, CCSD(T) [30, 31, 32], within the frozen core (fc) approximation. By using the correlation consistent basis sets cc-pVnZ basis set [33, 34, 35] both the Hartree-Fock self-consistent-field (HF-SCF) energies and the CCSD(T) correlation energies were extrapolated at the complete basis set (CBS) limit (at which the BSSE error vanishes), and then combined to correct for the error due to the basis-set truncation. CBS energies at HF-SCF level were evaluated using cc-pV$n$Z ($n$ = T, Q and 5) basis sets, and employing the $e^{-Cn}$ formula [36], while the CCSD(T) correlation contributions were computed using the $n^{-3}$ expression [37] with the cc-pVTZ and cc-pVQZ basis sets. Subsequent corrections due to core-valence (CV) correlation effects were calculated as the difference between CCSD(T) energies (in conjunction with the cc-pCVTZ basis set [38]) obtained by correlating all electrons and within the fc approximation. Inclusion of the ZPV corrections yielded the best estimate values for the binding energies reported in the present work, which therefore were then used to identify the most stable isomer for both OCS-HCCCCH and N$_2$O-HCCCCH complexes. The results for rotational constants, interaction and binding energies are summarized in Table 1.

For the most stable isomer of both OCS-HCCCCH and N$_2$O-HCCCCH complexes, we carried out additional geometry optimizations at both CCSD(T) and CCSD(T*)-F12c [39] levels of theory and in conjunction with the cc-pVTZ and cc-pV$n$Z-F12 [40] basis sets ($n$ = D, T), respectively; for the latter, both the appropriate auxiliary basis sets [41] and its complement auxiliary basis sets (CABS) [42] were used. Finally, for these two structures, within the framework of the vibrational second-order perturbation theory (VPT2 [43, 44]) we computed the vibrational corrections $\Delta B^i_{vib}$ to the corresponding equilibrium rotational constants $B^i_e$ by means of the following expression

$$\Delta B^i_{vib} = -\frac{1}{2}\sum_n \alpha^i_n \quad (2)$$

where the $\alpha^i_n$ are the vibration-rotation interaction constants ($i$ labels the inertial axis). These corrections were computed on the basis of the cubic force fields, which were calculated by using the maug-cc-pVTZ basis set and the B3PLYP functional (with the D3 corrections), in view of its good performance in modeling the anharmonic part of the potential [45], following the procedures established in previous works [46, 47].

We performed all DFT calculations employing the Gaussian suite of quantum chemical programs [48]; for both B3LYP and B2PLYP functionals, we employed the UltraFine grid available in Gaussian09 (corresponding to 99 radial and 590 angular points), because of its good results (see for example Refs. [18, 49]) for the calculations of anharmonic force field data. For the calculations carried out at coupled cluster levels of theory we used the MOLPRO program [50, 51] together with its appropriate software packages [52, 53, 54, 55, 56]. The optimized rotational parameters for the most stable isomers are given in Table 2.

Table 1. Theoretical molecular parameters for OCS-HCCCCH and $N_2$O-HCCCCH isomers optimized at DFT level of theory.[a]

|  | OCS-HCCCCH | | | | $N_2$O-HCCCCH | | |
|---|---|---|---|---|---|---|---|
|  | Isomer I (S-in) | Isomer II (O-in) | Isomer III T-shaped | Isomer IV Cross-shaped | Isomer I (N-in) | Isomer II (O-in) | Isomer III Cross-shaped |
| $A$[a] | 2897 | 4088 | 4255 | 2629 | 4461 | 4401 | 3262 |
| $B$[a] | 1277 | 1000 | 795 | 1243 | 1446 | 1470 | 1754 |
| $C$[a] | 886 | 803 | 670 | 1139 | 1092 | 1102 | 1394 |
| $BE_{(CP)}$[b] | -776 | -630 | -584 | -668 | -727 | -675 | -612 |
| $BE_{(CP+ZPV)}$[c] | -562 | -313 | -403 | -352 | -495 | -285 | -406 |
| $IE_{(CP)}$[d] | -780 | -633 | -584 | -668 | -727 | -679 | -612 |
| $BE_{(CBS)}$[e] | -682 | -560 | -504 | -567 | -700 | -611 | -570 |
| $BE_{(CBS+CV)}$[f] | -694 | -569 | -513 | -577 | -705 | -617 | -577 |
| $BE_{(CBS+CV+ZPV)}$[c] | -479 | -252 | -332 | -384 | -472 | -227 | -371 |

[a]Equilibrium rotational constants (in MHz) obtained from optimized geometries obtained with B2PLYP-D3 in conjunction with maug-cc-pVTZ basis set. Binding energies (BE) and interaction energies (IE) reported in cm$^{-1}$.

[b]Counterpoise-corrected binding energy computed at B2PLYP-D3/maug-cc-pVTZ level of theory.

[c]Binding energy including zero-point vibrational contribution evaluated at B2PLYP-D3/maug-cc-pVTZ level of theory.

[d]Counterpoise-corrected interaction energy computed at B2PLYP-D3/maug-cc-pVTZ level of theory.

[e]Binding energy obtained from extrapolated energies to the CBS limit at CCSD(T) level of theory (see text).

[g]Binding energy obtained from extrapolated energies to the CBS limit at CCSD(T) level of theory corrected by CV-effects (see text).

Table 2. Equilibrium rotational constants (in MHz) obtained at different coupled cluster levels of theory for the OCS-HCCCCH and $N_2O$-HCCCCH isomers experimentally observed in the present work.

|   | OCS-HCCCCH | | $N_2O$-HCCCCH | |
|---|---|---|---|---|
|   | Isomer I | | Isomer I | |
|   | CCSD(T)[a] | CCSD(T*)-F12c[b] | CCSD(T)[a] | CCSD(T*)-F12c[b] |
| A | 2856 | 2897 | 4487 | 4540 |
| B | 1205 | 1274 | 1395 | 1446 |
| C | 847 | 885 | 1064 | 1097 |

[a] Optimized geometries obtained at CCSD(T, fc) level of theory in conjunction with cc-pVTZ basis set.

[b] Optimized geometries obtained at CCSD(T*)-F12c level of theory in conjunction with VTZ-F12 basis set.

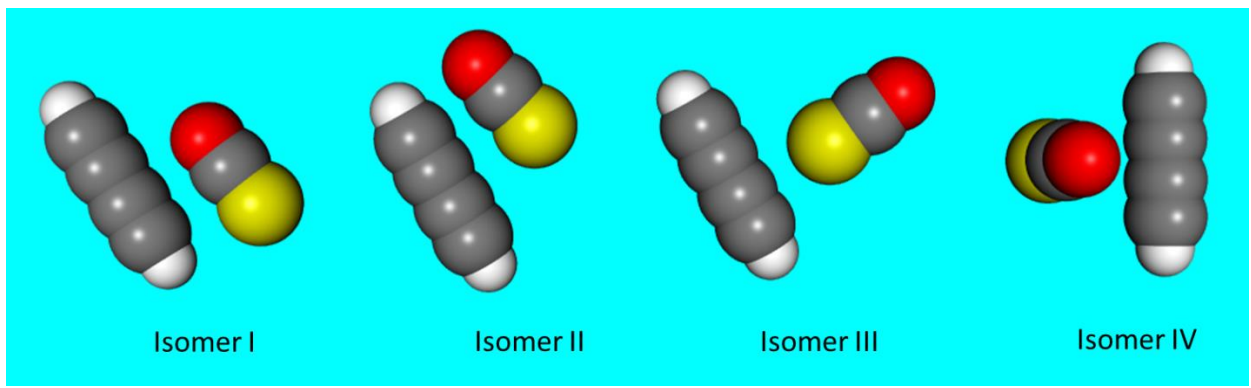

**Figure 1:** The four isomers of OCS-HCCCCH using DFT level of theory. Isomer I, studied in this work, is the lowest energy structure. See Table 1 for energy ordering of the isomers.

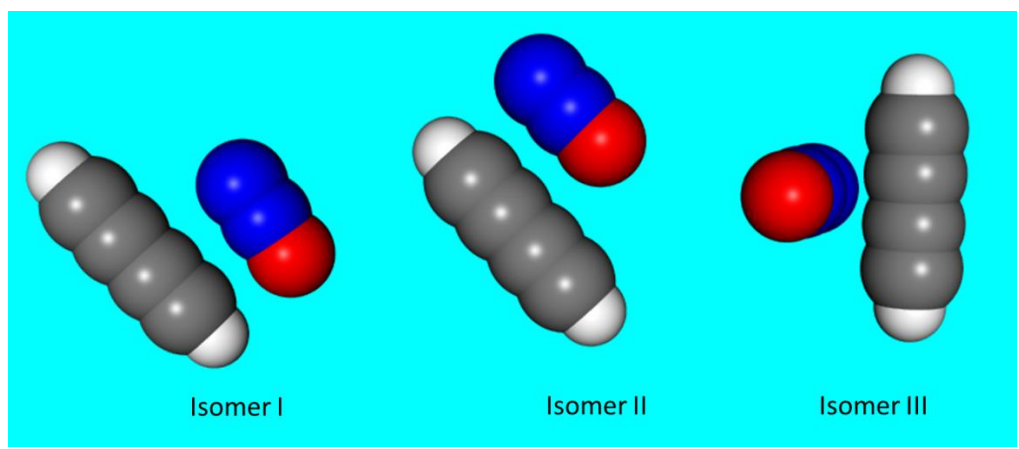

**Figure 2:** The three isomers of $N_2O$-HCCCCH using DFT level of theory. Isomer I, studied in this work, is the lowest energy structure. See Table 1 for energy ordering of the isomers.

## 2. Observed spectra

The spectra were recorded as described previously [57-59], using a pulsed supersonic slit jet apparatus, a diode laser for OCS-HCCCCH and a Daylight Solutions quantum cascade laser for $N_2O$-HCCCH. The expansion gas was a mixture of $N_2O$ (0.1%) and $C_4H_2$ or $C_4D_2$ (0.3%) in helium as a carrier gas with a jet backing pressure of 8 atmospheres. The lower concentration of $N_2O$ relative to HCCCCH was needed to minimize the formation of $N_2O$ clusters, which are known to absorb in the 2220 cm$^{-1}$ spectral region [3]. Similar conditions were used for OCS-HCCCCH and its deuterated isotopologue.

The diacetylene was synthesized by the procedure described in Ref. [60]. DCCCCD was obtained by mixing HCCCCH with a 1 N solution of NaOD in $D_2O$ as described by Etoh et al. [61]. The sample of diacetylene and diacetylene-$d_2$ thus prepared were purified by repeated distillation under vacuum and then stored at $LN_2$ temperature. The purity of the sample was checked using low resolution infrared spectroscopy. Spectral assignment and simulation were made using PGOPHER [62].

2.1. OCS-HCCCCH results

For all the minima of OCS-HCCCCH (listed in Table 1), the analysis of their structures (optimized at B2PLYP-D3/maug-cc-pVTZ level of theory) shows that there is an almost negligible variation in the intramolecular geometric parameters due to the complexation. Both the counterpoise corrected binding energies computed at DFT level of theory and those obtained from CCSD(T) extrapolated energies to the CBS limit identify isomer I (see Fig. 1) as the most stable form of OCS-HCCCCH (-2.22 kcal mol$^{-1}$ at B2PLYP level of theory, -1.95 kcal mol$^{-1}$ at CCSD(T) CBS limit). The inclusion of CV-effects is almost negligible, while the zero-point vibrational contributions reduce the binding energie (computed from extrapolated energies to the CBS limit at CCSD(T) level of theory) of isomer I to -1.37 kcal mol$^{-1}$. This isomer has a planar structure with nearly parallel side-by-side monomer units and the OCS S-atom on the inside (S-in). Furthermore, the calculated angle between the OCS monomer and the a inertial axis is 72°. Therefore the intramolecular fundamental in the region of the OCS monomer $\nu_1$ fundamental is an a/b-hybrid band, with a b-type transition moment which is larger than that for a-type transitions by a factor of ~3. Isomer II is also planar with a similar structure to isomer I, but with the OCS monomer unit in reverse orientation (O-in). Here, the calculated angle between the OCS monomer and the a inertial axis is 42°. This implies an a/b-hybrid band for the intramolecular fundamental in the $\nu_1$ region of the OCS with almost equal transition moments for the a- and b-type transitions. The expected rotational structure for isomer III is a nearly pure a-type band, as the calculated angle between the OCS monomer and the a inertial axis in only 12°. Finally, isomer IV is non-planar and would give rise to a mostly c-type band.

A segment of the experimental spectrum for OCS-HCCCCH, including the sharp Q-branch marked by an asterisk and a part of the R-branch, is shown in the second trace in Fig. 3. The corresponding segment for OCS-DCCCCD is illustrated in the bottom trace. With a 1:3 mixture of OCS and diacetylene we observe a relatively strong a/b-hybrid band, which could be simulated with a b-type transition moment about three times that of the a-type transition moment. This implies that the band observed is due to the most stable form. Further evidence for this assignment is given in section 3.

A total of 245 transitions involving rotational levels up to $J = 12$ and $k_a = 11$ were assigned for OCS-HCCCCH. Preliminary analysis of the spectrum confirmed that the upper state rotational levels with $k_a' > 6$ are perturbed. The assigned transitions were then used to obtain 90 ground state combination differences (GSCD). These were used in a frequency analysis to obtain five ground state parameters. The upper state parameters were subsequently obtained from a frequency analysis of the assigned transitions with $k_a' \leq 6$ while keeping the ground state parameters fixed. The frequency analysis for GSCD gave a weighted standard deviation of 0.00015 cm$^{-1}$ and that for the upper vibrational state was 0.00018 cm$^{-1}$. The parameters thus obtained are listed in the second column of Table 3. The top two traces in Figure 3 show a comparison of observed and simulated spectra based on these parameters.

Analysis of the band for OCS-DCCCCD was more straightforward, partly because the spectral region observed was more limited, to conserve the DCCCCD sample, and partly because the upper state rotational levels don't seem to be affected by perturbations. A total of 159 transitions involving levels up to $J = 12$ and $k_a = 8$ were assigned and analysed to obtain the parameters listed in column 3 of Table 3. The third and fourth traces in Fig. 3 show a comparison of simulated and observed spectra using these parameters. The assigned transitions are listed along with their corresponding residuals in Table A-1 for OCS-HCCCCH, including the perturbed transitions with $k_a' > 6$ which were given zero weight, and Table A-2 for OCS-DCCCCD.

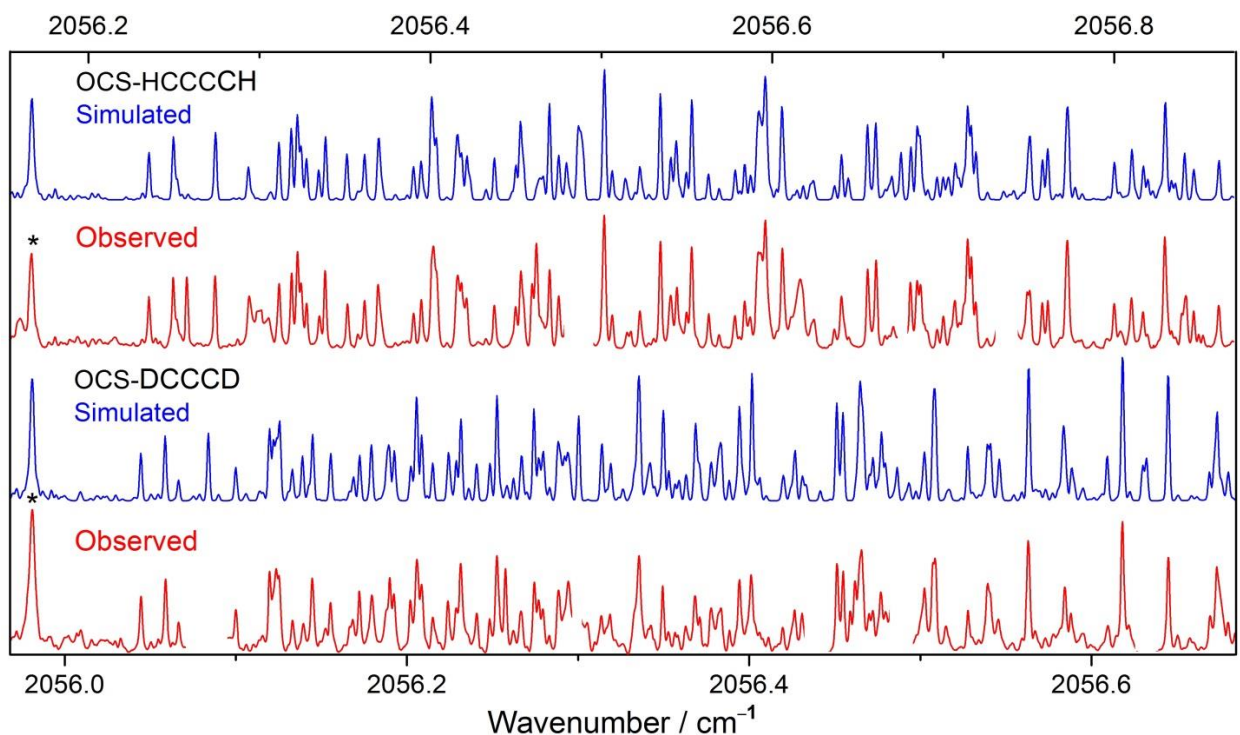

**Figure 3:** Observed and simulated spectra of OCS-HCCCCH and OCS-DCCCCD. The simulation is based on the fitted parameters of Table 3, an effective rotational temperature of 2.5 K, and an assumed Gaussian line width of 0.0018 cm$^{-1}$. Blank regions in the observed spectra are obscured by absorptions due to OCS monomer.

Table 3. Molecular parameters for OCS – HCCCCH.[a]

|  | This work | | CCSD(T*)-F12c | |
|---|---|---|---|---|
|  | OCS – HCCCCH | OCS – DCCCCD | OCS – HCCCCH | OCS – DCCCCD |
| $\nu_0$ / cm$^{-1}$ | 2056.16809(3) | 2055.98241(5) | | |
| $A'$ / MHz | 2891.907(68) | 2692.82(28) | | |
| $B'$ / MHz | 1242.930(34) | 1205.53(12) | | |
| $C'$ / MHz | 867.841(16) | 831.664(79) | | |
| $D_K'$ / kHz | 6.3(22)[c] | | | |
| $D_{JK}'$ / kHz | 5.7(19)[c] | 15.6(24) | | |
| $A''$ / MHz | 2892.78(15)[b] | 2693.92(28) | 2882.07[e] | 2684.26[e] |
| $B''$ / MHz | 1244.416(74)[b] | 1206.46(12) | 1261.10[e] | 1223.09[e] |
| $C''$ / MHz | 869.008(55)[b] | 832.221(83) | 876.38[e] | 839.43[e] |
| $D_K''$ / kHz | 6.3(22)[b] | | | |
| $D_{JK}''$ / kHz | 5.7(19)[b] | 12.8(24) | | |
| $\Delta$ / a.m.u Å$^2$ | 0.78[d] | 0.77[d] | | |

[a] Uncertainties (1σ) in parentheses are in units of the last quoted digit.
[b] Ground state parameters were obtained from the lower state combination differences. These were held fixed during the frequency analysis of the upper state rotational levels.
[c] Excited state quartic parameters were held fixed at their ground state values.
[d] Inertial defect, $\Delta = I_c - I_a - I_b$.
[e] Equilibrium rotational constants at CCSD(T*)-F12c/VTZ-F12 level of theory augmented by vibrational corrections computed at B3LYP-D3 level (see text).

## 2.2. N$_2$O-HCCCCH results

Figure 2 illustrates the three possible forms of N$_2$O-HCCCCH identified in the present work. For all the structures (obtained at B2PLYP-D3/maug-cc-pVTZ level of theory) the complexation leads to negligible modifications in the intramolecular geometric parameters, as found for the isomers of OCS-HCCCCH. All the calculations carried out in the present work (see Table 1) identify isomer I (N-in) as the most stable form. Its binding energy, computed from extrapolated energies to the CBS limit at CCSD(T) level of theory corrected by CV-effects and taking into account zero-point vibrational contributions, is -1.35 kcal mol$^{-1}$. This isomer has a planar structure with two nearly parallel monomer units. Here, the calculated angle between the N$_2$O monomer and the a principal axis is 56°, implying that the intramolecular fundamental in the region of the N$_2$O monomer $\nu_1$ fundamental is an a/b-hybrid band with a b-type transition moment about 1.5 times that for the a-type transitions. The higher energy isomer (O-in) has a very similar structure with the N$_2$O monomer making an angle of 53° with the a axis, only 3° less than isomer I; thus the expected intramolecular bands for the two isomers have very similar rotational structure. Isomer III is non-planar with the N$_2$O monomer axis perpendicular to the ab plane, resulting in a purely c-type band.

A segment of the experimental spectrum for N$_2$O-HCCCCH is shown in the second trace of Fig. 4. Unlike OCS-HCCCCH, the Q-branch, marked by an asterisk, shows much more structure. The corresponding segment for N$_2$O-DCCCCD is illustrated in the bottom trace. Again, we observe a relatively strong a/b-hybrid band which is simulated with the b-type transition moment ~1.5 times that of the a-type transition moment. Although the simulated spectrum with this ratio of the transition moments (1.5) gives better overall intensity agreement with the experimental spectrum than b-type/a-type = 1.3 as calculated for isomer II, this alone is insufficient to conclusively assign the observed band to isomer I. Further evidence in support of assignment of the observed band to isomer I is given in section 3.

Analysis of the bands for $N_2O$-HCCCCH and $N_2O$-DCCCCD was straightforward. A total of 166 transitions involving levels up to $J = 11$ and $k_a = 8$ for $N_2O$-HCCCCH and 121 transitions involving levels up to $J = 10$ and $k_a = 5$ were assigned and analysed to obtain the parameters listed in columns 3 and 4 of Table 4. The weighted standard deviation of the fit for $N_2O$-HCCCCH was 0.00023 cm$^{-1}$ and that for $N_2O$-DCCCCD was 0.00024 cm$^{-1}$. The first and third traces in Fig. 4 show the simulated spectra using these parameters. The assigned transitions along with their corresponding residuals are listed in Table A-3 for $N_2O$-HCCCCH and Table A-4 for $N_2O$-DCCCCD.

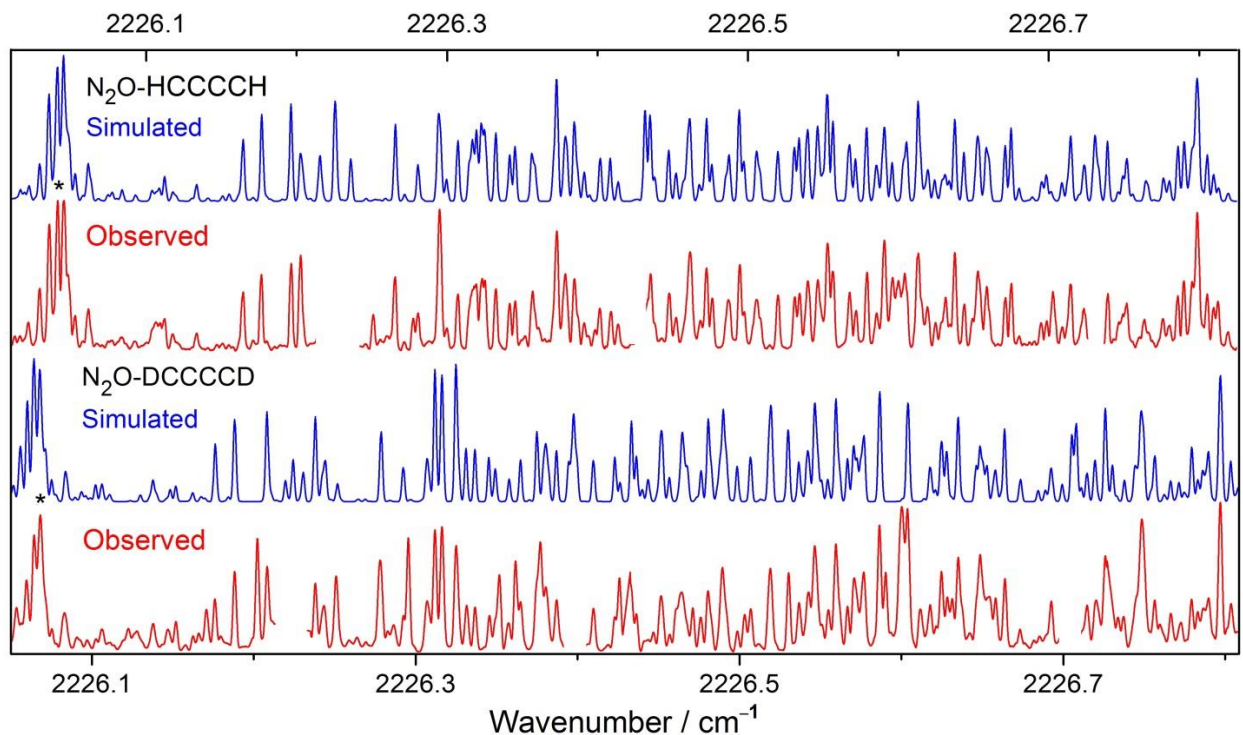

**Figure 4: Observed and simulated spectra of $N_2O$-HCCCCH and $N_2O$-DCCCCD. The simulation is based on the fitted parameters of Table 4, an effective rotational temperature of 2.5 K, and an assumed Gaussian line width of 0.0022 cm$^{-1}$. Blank regions in the observed spectra are obscured by absorptions due to $N_2O$ monomer.**

Table 4. Molecular parameters for N$_2$O – HCCCCH.[a]

|  | This work | | CCSD(T*)-F12c | |
| --- | --- | --- | --- | --- |
|  | N$_2$O – HCCCCH | N$_2$O – DCCCCD | N$_2$O – HCCCCH | N$_2$O – DCCCCD |
| $\nu_0$ / cm$^{-1}$ | 2226.05079(5) | 2226.07274(5) |  |  |
| $A'$ / MHz | 4479.78(35) | 4117.27(20) |  |  |
| $B'$ / MHz | 1409.16(12) | 1358.51(12) |  |  |
| $C'$ / MHz | 1069.838(81) | 1019.606(90) |  |  |
| $D_K'$ / kHz | 82.9(77) |  |  |  |
| $D_{JK}'$ / kHz | -12.2(25) |  |  |  |
| $A''$ / MHz | 4498.57(28) | 4134.60(52) | 4485.29[c] | 4117.35[c] |
| $B''$ / MHz | 1409.03(12) | 1358.33(12) | 1434.99[c] | 1384.70[c] |
| $C''$ / MHz | 1070.783(76) | 1020.552(99) | 1085.21[c] | 1034.36[c] |
| $D_K''$ / kHz | 91.8(51) | 93(28) |  |  |
| $\Delta$ / a.m.u Å$^2$ | 0.96[b] | 0.91[b] |  |  |

[a] Uncertainties (1σ) in parentheses are in units of the last quoted digit.
[b] Inertial defect, $\Delta = I_c - I_a - I_b$.
[c] Equilibrium rotational constants at CCSD(T*)-F12c/VTZ-F12 level of theory augmented by vibrational corrections computed at B3LYP-D3 level (see text).

## 3. Discussion and conclusions

Comparison of the theoretical rotational constants for OCS-HCCCCH at all levels of theory, columns 2-5 in Table 1 and columns 2-3 in Table 2, with the experimental rotational constants, column 2 in Table 3, clearly indicates that the observed dimer corresponds to the lowest energy isomer identified in the present work. Moreover, the best agreement for both the dimers is obtained at the highest level of theory, CCSD(T*)-F12c/VTZ-F12. The results for the deuterated isotopologue, columns 3 and 4 in Table 3, provide further support for this assignment. This, together with the expectation that the jet conditions employed in this work favor the formation of the lowest energy structure, leaves very little doubt that isomer I in Fig. 1 corresponds to the species observed.

Small and positive values for the inertial defects (Table 3) indicate that observed dimer is indeed planar. Positive contributions to the inertial defect result from Coriolis interactions associated with in-plane bending motions; in isomer I of OCS-HCCCCH these are larger than negative contributions from out-of-plane vibrations [63].

All of the arguments made above in support of isomer I of OCS-HCCCCH being the carrier of the observed band in the region of the $\nu_1$ fundamental of OCS can similarly be made for the most stable isomer of $N_2O$-HCCCCH. Therefore, the carrier of the intramolecular band in the region of the $\nu_1$ fundamental of $N_2O$ is isomer I in Fig. 2. Again, the small and positive values for the inertial defects (Table 4) indicate that the observed dimer is planar. The slight increase in the magnitude of inertial defect, from 0.78 a.m.u. Å$^2$ for OCS-HCCCCH to 0.96 a.m.u. Å$^2$, is consistent with the fact that $N_2O$-HCCCCH is a lighter dimer.

The theoretical intermolecular distances for OCS-HCCCCH (S-in) and $N_2O$-HCCCCH (N-in) with the experimental vibrational shifts are listed in Table 5. In the case of OCS-HCCCCH, these are compared with the corresponding values for the two observed isomers of OCS-HCCH [3]. As can be seen, the intermolecular distance of R = 3.69 Å for OCS-HCCCCH (S-in) is comparable to that for the near-parallel isomer of OCS-HCCH (S-in, R = 3.61 Å) and expectedly much shorter than that for the T-shaped structure (R = 4.62). In the case of $N_2O$-

HCCCCH (N-in), the calculated intermolecular distance is for 3.43 Å, somewhat larger than R = 3.30 Å for $N_2O$-HCCH (O-in). The larger R for $N_2O$-HCCCCH (N-in) may be explained by the combination of the Van der Waals radii, 1.55 Å for the oxygen atom and 1.50 Å for nitrogen [64], and the fact that the center of mass is shifted by 0.07 Å from the central nitrogen atom toward the oxygen atom in the $N_2O$ monomer unit.

As always, the vibrational shifts are difficult to interpret. However, we note that the shifts for both diacetylene- and acetylene-containing species, as shown in Table 5, are consistently to lower frequency for dimers containing OCS and to higher frequency for dimers containing $N_2O$.

Table 5. Comparison of intermolecular distance and vibrational shift between acetylene and diacetylene containing dimers

|  | OCS-HCCCCH Near-parallel (S-in) | OCS-HCCH Near-parallel (S-in) | OCS-HCCH T-shaped[a] | $N_2O$-HCCCCH Near-parallel (N-in)[b] | $N_2O$-HCCH Near-parallel (O-in) |
|---|---|---|---|---|---|
| Intermolecular distance R (Å) | 3.69 | 3.61 | 4.62 | 3.43 | 3.30 |
| Vibrational Shift ($cm^{-1}$) | -6.085 | -0.286 | -5.688 | +2.219 | +5.346 |

[a] The calculated intermolecular distance for T-shaped OCS-HCCCCH is 4.52 Å.

[b] For comparison the calculated intermolecular distance for $N_2O$-HCCCCH (O-in) is 3.60 Å.

In summary, we have observed and analyzed spectra of the most stable isomers of OCS-HCCCCH and $N_2O$-HCCCCH using the vibrational fundamentals of OCS and $N_2O$ monomers in the 4 micron region as IR chromophores. Both dimers were found to have planar structure with nearly parallel monomer units. A combination of experimental evidence and theoretical calculations, at several different levels of theory, was used to identify the carriers of the infrared bands. Theoretical calculations predicted four isomers for OCS-HCCCCH and three for $N_2O$-HCCCCH, and the observed structures were assigned to the lowest energy structures on the potential energy surfaces. The observed spectra for OCS-DCCCCD and $N_2O$-DCCCCD are entirely consistent with the normal isotopologues. Given the similarities between the most stable structures for OCS-HCCCCH with OCS-HCCH and for $N_2O$-HCCCCH with $N_2O$-HCCH, we conclude that diacetylene closely parallels acetylene upon complexation with OCS or $N_2O$.

**Acknowledgements**

The financial support of the Natural Sciences and Engineering Research Council of Canada is gratefully acknowledged. The High Performance Computing department of the CINECA Supercomputer Centre (grant no. HP10C8R8EI) and the SCSCF ("Sistema per il Calcolo Scientifico di Ca′ Foscari") facility are gratefully acknowledged for the utilization of computer resources. A.C.P. gratefully acknowledge the financial support of Università Ca' Foscari Venezia (ADiR funds).

**Appendix A. Supplementary data**

Supplementary data associated with this article can be found in the online version of the manuscript.

Table A-1. Observed and calculated transitions for OCS-HCCCCH in the region of the $\nu_1$ fundamental band of OCS (units of cm$^{-1}$).

```
*************************************************************************
 J' Ka' Kc'  J" Ka" Kc"   Observed    Calculated  Obs-Calc  Weight
*************************************************************************
 9   5   5   10  6   4   2054.78574  2054.78454   0.00119    0.25
 9   5   4   10  6   5   2054.78574  2054.78566   0.00007    0.25
 7   6   1    8  7   2   2054.80586  2054.80560   0.00025    0.25
 7   6   2    8  7   1   2054.80586  2054.80560   0.00025    0.25
 9   4   6   10  5   5   2054.90163  2054.90135   0.00027    0.25
 7   5   3    8  6   2   2054.92716  2054.92732  -0.00015    0.25
 7   5   2    8  6   3   2054.92716  2054.92740  -0.00023    0.25
 8   4   4    9  5   5   2054.98389  2054.98370   0.00018    1.00
 6   5   2    7  6   1   2054.99872  2054.99886  -0.00014    0.25
 6   5   1    7  6   2   2054.99872  2054.99887  -0.00015    0.25
 6   5   1    7  6   2   2054.99872  2054.99887  -0.00015    0.25
 6   5   2    7  6   1   2054.99872  2054.99886  -0.00014    0.25
 7   4   4    8  5   3   2055.04882  2055.04872   0.00010    1.00
 7   4   3    8  5   4   2055.05140  2055.05131   0.00009    1.00
 5   5   1    6  6   0   2055.07068  2055.07044   0.00023    0.25
 5   5   0    6  6   1   2055.07068  2055.07044   0.00023    0.25
 6   4   2    7  5   3   2055.12095  2055.12124  -0.00028    0.25
 6   4   3    7  5   2   2055.12095  2055.12052   0.00042    0.25
 8   3   5    9  4   6   2055.14721  2055.14712   0.00008    1.00
 7   3   5    8  4   4   2055.15611  2055.15621  -0.00009    1.00
 5   4   1    6  5   2   2055.19192  2055.19223  -0.00031    0.25
 5   4   2    6  5   1   2055.19192  2055.19209  -0.00017    0.25
 7   3   4    8  4   5   2055.19521  2055.19534  -0.00013    1.00
 6   3   4    7  4   3   2055.23641  2055.23653  -0.00011    1.00
 6   3   3    7  4   4   2055.25275  2055.25291  -0.00015    1.00
 4   4   0    5  5   1   2055.26320  2055.26362  -0.00042    0.25
```

| | | | | | | | | | |
|---|---|---|---|---|---|---|---|---|---|
| 4  | 4 | 1  | 5  | 5 | 0  | 2055.26320 | 2055.26361 | -0.00040 | 0.25 |
| 6  | 2 | 5  | 7  | 3 | 4  | 2055.29749 | 2055.29772 | -0.00022 | 1.00 |
| 5  | 3 | 3  | 6  | 4 | 2  | 2055.31161 | 2055.31191 | -0.00029 | 1.00 |
| 5  | 3 | 2  | 6  | 4 | 3  | 2055.31750 | 2055.31748 | 0.00001  | 1.00 |
| 8  | 2 | 6  | 9  | 3 | 7  | 2055.36010 | 2055.36014 | -0.00003 | 1.00 |
| 11 | 3 | 9  | 12 | 4 | 8  | 2055.36796 | 2055.36715 | 0.00080  | 1.00 |
| 5  | 1 | 5  | 6  | 2 | 4  | 2055.37112 | 2055.37133 | -0.00020 | 1.00 |
| 11 | 3 | 8  | 12 | 2 | 11 | 2055.37506 | 2055.37462 | 0.00043  | 1.00 |
| 7  | 2 | 5  | 8  | 3 | 6  | 2055.39129 | 2055.39159 | -0.00029 | 1.00 |
| 5  | 2 | 4  | 6  | 3 | 3  | 2055.40216 | 2055.40241 | -0.00025 | 1.00 |
| 11 | 0 | 11 | 12 | 1 | 12 | 2055.43818 | 2055.43790 | 0.00027  | 0.25 |
| 11 | 1 | 11 | 12 | 0 | 12 | 2055.43818 | 2055.43814 | 0.00003  | 0.25 |
| 10 | 2 | 9  | 11 | 3 | 8  | 2055.44544 | 2055.44539 | 0.00004  | 1.00 |
| 3  | 3 | 0  | 4  | 4 | 1  | 2055.45660 | 2055.45688 | -0.00028 | 0.25 |
| 3  | 3 | 1  | 4  | 4 | 0  | 2055.45660 | 2055.45668 | -0.00008 | 0.25 |
| 5  | 2 | 3  | 6  | 3 | 4  | 2055.47156 | 2055.47185 | -0.00029 | 1.00 |
| 9  | 1 | 8  | 10 | 2 | 9  | 2055.48121 | 2055.48133 | -0.00011 | 1.00 |
| 4  | 2 | 3  | 5  | 3 | 2  | 2055.49264 | 2055.49289 | -0.00025 | 1.00 |
| 10 | 1 | 10 | 11 | 0 | 11 | 2055.49706 | 2055.49710 | -0.00004 | 0.25 |
| 10 | 0 | 10 | 11 | 1 | 11 | 2055.49706 | 2055.49658 | 0.00048  | 0.25 |
| 9  | 2 | 8  | 10 | 1 | 9  | 2055.51063 | 2055.51052 | 0.00011  | 1.00 |
| 4  | 2 | 2  | 5  | 3 | 3  | 2055.52359 | 2055.52391 | -0.00031 | 1.00 |
| 8  | 1 | 7  | 9  | 2 | 8  | 2055.52909 | 2055.52920 | -0.00010 | 1.00 |
| 8  | 2 | 7  | 9  | 1 | 8  | 2055.57994 | 2055.57996 | -0.00002 | 1.00 |
| 3  | 2 | 1  | 4  | 3 | 2  | 2055.58407 | 2055.58425 | -0.00017 | 1.00 |
| 8  | 1 | 8  | 9  | 0 | 9  | 2055.61533 | 2055.61541 | -0.00007 | 1.00 |
| 8  | 1 | 8  | 9  | 0 | 9  | 2055.61541 | 2055.61541 | 0.00000  | 1.00 |
| 3  | 1 | 3  | 4  | 2 | 2  | 2055.63864 | 2055.63860 | 0.00004  | 1.00 |
| 2  | 2 | 1  | 3  | 3 | 0  | 2055.64887 | 2055.64891 | -0.00003 | 1.00 |
| 7  | 2 | 6  | 8  | 1 | 7  | 2055.65504 | 2055.65542 | -0.00037 | 1.00 |
| 7  | 0 | 7  | 8  | 1 | 8  | 2055.67033 | 2055.67035 | -0.00002 | 1.00 |
| 7  | 1 | 7  | 8  | 0 | 8  | 2055.67517 | 2055.67520 | -0.00003 | 1.00 |
| 10 | 2 | 9  | 10 | 3 | 8 | 2055.70243 | 2055.70254 | -0.00011 | 1.00 |
| 10 | 3 | 8  | 10 | 4 | 7 | 2055.70452 | 2055.70451 |  0.00000 | 1.00 |
|  9 | 3 | 7  |  9 | 4 | 6 | 2055.72130 | 2055.72148 | -0.00017 | 1.00 |
|  6 | 0 | 6  |  7 | 1 | 7 | 2055.72639 | 2055.72645 | -0.00006 | 1.00 |
|  7 | 3 | 4  |  7 | 4 | 3 | 2055.76826 | 2055.76864 | -0.00038 | 1.00 |
|  8 | 0 | 8  |  8 | 1 | 7 | 2055.77681 | 2055.77681 | -0.00000 | 1.00 |
|  5 | 0 | 5  |  6 | 1 | 6 | 2055.78047 | 2055.78064 | -0.00017 | 1.00 |
| 10 | 1 | 9  | 10 | 2 | 8 | 2055.78362 | 2055.78354 |  0.00008 | 1.00 |
|  2 | 1 | 1  |  3 | 2 | 2 | 2055.79137 | 2055.79139 | -0.00002 | 1.00 |
|  5 | 1 | 5  |  6 | 0 | 6 | 2055.79864 | 2055.79879 | -0.00014 | 1.00 |
|  7 | 2 | 6  |  7 | 3 | 5 | 2055.80888 | 2055.80889 | -0.00000 | 1.00 |
|  5 | 2 | 4  |  6 | 1 | 5 | 2055.82692 | 2055.82657 |  0.00035 | 1.00 |
|  1 | 1 | 0  |  2 | 2 | 1 | 2055.84992 | 2055.84955 |  0.00036 | 1.00 |
|  4 | 1 | 4  |  5 | 0 | 5 | 2055.86473 | 2055.86454 |  0.00018 | 1.00 |
|  4 | 2 | 2  |  4 | 3 | 1 | 2055.88015 | 2055.87996 |  0.00019 | 1.00 |
|  5 | 2 | 3  |  5 | 3 | 2 | 2055.89600 | 2055.89603 | -0.00003 | 1.00 |
|  2 | 0 | 2  |  3 | 1 | 3 | 2055.93210 | 2055.93187 |  0.00023 | 1.00 |
|  3 | 1 | 3  |  3 | 2 | 2 | 2055.94540 | 2055.94520 |  0.00019 | 1.00 |
|  7 | 1 | 6  |  7 | 2 | 5 | 2055.96164 | 2055.96176 | -0.00011 | 1.00 |
|  5 | 0 | 5  |  5 | 1 | 4 | 2055.97204 | 2055.97218 | -0.00014 | 1.00 |
|  1 | 0 | 1  |  2 | 1 | 2 | 2055.98462 | 2055.98455 |  0.00007 | 1.00 |
|  6 | 1 | 5  |  6 | 2 | 4 | 2055.99379 | 2055.99393 | -0.00014 | 1.00 |
|  2 | 1 | 1  |  2 | 2 | 0 | 2056.00104 | 2056.00097 |  0.00006 | 1.00 |
|  2 | 1 | 2  |  3 | 0 | 3 | 2056.00820 | 2056.00808 |  0.00012 | 1.00 |
|  4 | 1 | 3  |  4 | 2 | 2 | 2056.01557 | 2056.01574 | -0.00017 | 1.00 |
|  4 | 0 | 4  |  4 | 1 | 3 | 2056.02294 | 2056.02297 | -0.00003 | 1.00 |
|  0 | 0 | 0  |  1 | 1 | 1 | 2056.04269 | 2056.04261 |  0.00008 | 1.00 |
|  3 | 0 | 3  |  3 | 1 | 2 | 2056.06072 | 2056.06081 | -0.00009 | 1.00 |
|  1 | 0 | 1  |  1 | 1 | 0 | 2056.10038 | 2056.10049 | -0.00011 | 1.00 |
|  1 | 1 | 0  |  1 | 0 | 1 | 2056.23532 | 2056.23551 | -0.00019 | 1.00 |
|  2 | 1 | 1  |  2 | 0 | 2 | 2056.24956 | 2056.24975 | -0.00019 | 1.00 |
|  3 | 1 | 2  |  3 | 0 | 3 | 2056.27402 | 2056.27431 | -0.00029 | 1.00 |
| 1 | 1 | 1 | 0 | 0 | 0 | 2056.29430 | 2056.29350 | 0.00079 | 1.00 |
| 1 | 1 | 1 | 0 | 0 | 0 | 2056.29430 | 2056.29350 | 0.00079 | 1.00 |
| 1 | 1 | 1 | 0 | 0 | 0 | 2056.29430 | 2056.29350 | 0.00079 | 1.00 |
| 4 | 1 | 3 | 4 | 0 | 4 | 2056.31152 | 2056.31146 | 0.00005 | 1.00 |
| 4 | 2 | 2 | 4 | 1 | 3 | 2056.31873 | 2056.31866 | 0.00006 | 1.00 |
| 3 | 2 | 1 | 3 | 1 | 2 | 2056.32431 | 2056.32431 | 0.00000 | 1.00 |
| 3 | 0 | 3 | 2 | 1 | 2 | 2056.32748 | 2056.32735 | 0.00013 | 1.00 |
| 2 | 2 | 0 | 2 | 1 | 1 | 2056.33499 | 2056.33473 | 0.00026 | 1.00 |
| 6 | 2 | 4 | 6 | 1 | 5 | 2056.33845 | 2056.33845 | -0.00000 | 1.00 |
| 2 | 1 | 2 | 1 | 0 | 1 | 2056.35146 | 2056.35131 | 0.00015 | 1.00 |
| 5 | 1 | 4 | 5 | 0 | 5 | 2056.36143 | 2056.36140 | 0.00003 | 1.00 |
| 7 | 2 | 5 | 7 | 1 | 6 | 2056.36954 | 2056.36939 | 0.00015 | 1.00 |
| 3 | 2 | 2 | 3 | 1 | 3 | 2056.39009 | 2056.39002 | 0.00007 | 1.00 |
| 8 | 3 | 5 | 8 | 2 | 6 | 2056.39469 | 2056.39451 | 0.00018 | 1.00 |
| 4 | 0 | 4 | 3 | 0 | 3 | 2056.43246 | 2056.43235 | 0.00010 | 1.00 |
| 5 | 3 | 2 | 5 | 2 | 3 | 2056.43745 | 2056.43758 | -0.00012 | 1.00 |
| 5 | 2 | 4 | 5 | 1 | 5 | 2056.44988 | 2056.44984 | 0.00004 | 1.00 |
| 5 | 0 | 5 | 4 | 1 | 4 | 2056.46968 | 2056.46959 | 0.00009 | 1.00 |
| 5 | 1 | 5 | 4 | 0 | 4 | 2056.50167 | 2056.50172 | -0.00005 | 1.00 |
| 6 | 1 | 5 | 5 | 2 | 4 | 2056.50626 | 2056.50648 | -0.00021 | 1.00 |
| 7 | 3 | 5 | 7 | 2 | 6 | 2056.52256 | 2056.52255 | 0.00000 | 1.00 |
| 6 | 0 | 6 | 5 | 1 | 5 | 2056.53447 | 2056.53447 | 0.00000 | 1.00 |
| 8 | 4 | 4 | 8 | 3 | 5 | 2056.54053 | 2056.54061 | -0.00007 | 1.00 |
| 8 | 4 | 4 | 8 | 3 | 5 | 2056.54053 | 2056.54061 | -0.00007 | 1.00 |
| 8 | 4 | 4 | 8 | 3 | 5 | 2056.54053 | 2056.54061 | -0.00007 | 1.00 |
| 3 | 2 | 2 | 2 | 1 | 1 | 2056.54412 | 2056.54405 | 0.00007 | 1.00 |
| 3 | 2 | 2 | 2 | 1 | 1 | 2056.54412 | 2056.54405 | 0.00007 | 1.00 |
| 3 | 2 | 2 | 2 | 1 | 1 | 2056.54412 | 2056.54405 | 0.00007 | 1.00 |
| 8 | 3 | 6 | 8 | 2 | 7 | 2056.54984 | 2056.54983 | 0.00000 | 1.00 |
| 8 | 3 | 6 | 8 | 2 | 7 | 2056.54984 | 2056.54983 | 0.00000 | 1.00 |
| 8 | 3 | 6 | 8 | 2 | 7 | 2056.54984 | 2056.54983 | 0.00000 | 1.00 |
| 6 | 1 | 6 | 5 | 0 | 5 | 2056.55276 | 2056.55260 | 0.00015 | 1.00 |
| 7 | 4 | 3 | 7 | 3 | 4 | 2056.56282 | 2056.56280 | 0.00002 | 1.00 |
| 8 | 2 | 6 | 7 | 3 | 5 | 2056.56882 | 056.569097 | -0.00026 | 1.00 |
| 6 | 4 | 2 | 6 | 3 | 3 | 2056.57806 | 2056.57818 | -0.00011 | 1.00 |
| 5 | 4 | 1 | 5 | 3 | 2 | 2056.58687 | 2056.58725 | -0.00038 | 1.00 |
| 7 | 1 | 7 | 6 | 0 | 6 | 2056.60589 | 2056.60576 | 0.00012 | 1.00 |
| 9 | 2 | 8 | 9 | 1 | 9 | 2056.63641 | 2056.63631 | 0.00009 | 1.00 |
| 5 | 2 | 4 | 4 | 1 | 3 | 2056.64049 | 2056.64040 | 0.00008 | 1.00 |
| 8 | 0 | 8 | 7 | 1 | 7 | 2056.65586 | 2056.65583 | 0.00002 | 1.00 |
| 8 | 1 | 8 | 7 | 0 | 7 | 2056.66063 | 2056.66067 | -0.00004 | 1.00 |
| 6 | 2 | 5 | 5 | 1 | 4 | 2056.68088 | 2056.68096 | -0.00008 | 1.00 |
| 3 | 3 | 1 | 2 | 2 | 0 | 2056.68476 | 2056.68458 | 0.00017 | 1.00 |
| 3 | 3 | 1 | 2 | 2 | 0 | 2056.68476 | 2056.68458 | 0.00017 | 1.00 |
| 3 | 3 | 0 | 2 | 2 | 1 | 2056.68655 | 2056.68667 | -0.00012 | 1.00 |
| 3 | 3 | 0 | 2 | 2 | 1 | 2056.68655 | 2056.68667 | -0.00012 | 1.00 |
| 4 | 2 | 2 | 3 | 1 | 3 | 2056.69637 | 2056.69623 | 0.00013 | 1.00 |
| 8 | 5 | 3 | 8 | 4 | 4 | 2056.70012 | 2056.70008 | 0.00003 | 1.00 |
| 9 | 0 | 9 | 8 | 1 | 8 | 2056.71423 | 2056.71428 | -0.00005 | 1.00 |
| 9 | 1 | 9 | 8 | 0 | 8 | 2056.71641 | 2056.71665 | -0.00023 | 1.00 |
| 7 | 2 | 6 | 6 | 1 | 5 | 2056.71895 | 2056.71918 | -0.00023 | 1.00 |
| 8 | 2 | 7 | 7 | 1 | 6 | 2056.75793 | 2056.75804 | -0.00010 | 1.00 |
| 4 | 3 | 1 | 3 | 2 | 2 | 2056.76120 | 2056.76120 | -0.00000 | 1.00 |
| 10 | 0 | 10 | 9 | 1 | 9 | 2056.77257 | 2056.77205 | 0.00052 | 0.25 |
| 10 | 1 | 10 | 9 | 0 | 9 | 2056.77257 | 2056.77317 | -0.00060 | 0.25 |
| 9 | 2 | 8 | 8 | 1 | 7 | 2056.80010 | 2056.80009 | 0.00000 | 1.00 |
| 5 | 3 | 3 | 4 | 2 | 2 | 2056.81014 | 2056.81019 | -0.00005 | 1.00 |
| 10 | 1 | 9 | 9 | 2 | 8 | 2056.81691 | 2056.81725 | -0.00034 | 1.00 |
| 11 | 1 | 11 | 10 | 0 | 10 | 2056.82965 | 2056.82996 | -0.00030 | 0.25 |
| 11 | 0 | 11 | 10 | 1 | 10 | 2056.82965 | 2056.82943 | 0.00021 | 0.25 |
| 10 | 2 | 9 | 9 | 1 | 8 | 2056.84647 | 2056.84642 | 0.00004 | 1.00 |
| 6 | 3 | 4 | 5 | 2 | 3 | 2056.86120 | 2056.86123 | -0.00003 | 1.00 |
| 12 | 1 | 12 | 11 | 0 | 11 | 2056.88645 | 2056.88682 | -0.00037 | 0.25 |
| 12 | 0 | 12 | 11 | 1 | 11 | 2056.88645 | 2056.88658 | -0.00013 | 0.25 |

| | | | | | | | | |
|---|---|---|---|---|---|---|---|---|
| 11 | 2 | 10 | 10 | 1 | 9 | 2056.89657 | 2056.89662 | -0.00004 | 1.00 |
| 7 | 3 | 5 | 6 | 2 | 4 | 2056.90371 | 2056.90373 | -0.00002 | 1.00 |
| 6 | 3 | 3 | 5 | 2 | 4 | 2056.93080 | 2056.93073 | 0.00007 | 1.00 |
| 8 | 3 | 6 | 7 | 2 | 5 | 2056.93892 | 2056.93894 | -0.00002 | 1.00 |
| 12 | 3 | 9 | 11 | 2 | 10 | 2056.94087 | 2056.94120 | -0.00033 | 1.00 |
| 9 | 3 | 7 | 8 | 2 | 6 | 2056.96887 | 2056.96887 | -0.00000 | 1.00 |
| 10 | 3 | 8 | 9 | 2 | 7 | 2056.99650 | 2056.99621 | 0.00029 | 1.00 |
| 14 | 0 | 14 | 13 | 1 | 13 | 2057.00025 | 2057.00048 | -0.00022 | 0.25 |
| 14 | 1 | 14 | 13 | 0 | 13 | 2057.00025 | 2057.00053 | -0.00027 | 0.25 |
| 13 | 2 | 12 | 12 | 3 | 9 | 2057.00415 | 2057.00426 | -0.00011 | 1.00 |
| 6 | 4 | 3 | 5 | 3 | 2 | 2057.01575 | 2057.01584 | -0.00008 | 1.00 |
| 6 | 4 | 2 | 5 | 3 | 3 | 2057.02139 | 2057.02142 | -0.00003 | 1.00 |
| 11 | 3 | 9 | 10 | 2 | 8 | 2057.02495 | 2057.02417 | 0.00077 | 1.00 |
| 5 | 5 | 0 | 4 | 4 | 1 | 2057.07104 | 2057.07097 | 0.00006 | 0.25 |
| 5 | 5 | 1 | 4 | 4 | 0 | 2057.07104 | 2057.07096 | 0.00008 | 0.25 |
| 7 | 4 | 4 | 6 | 3 | 3 | 2057.07910 | 2057.07921 | -0.00010 | 1.00 |
| 7 | 4 | 3 | 6 | 3 | 4 | 2057.09541 | 2057.09562 | -0.00020 | 1.00 |
| 16 | 2 | 14 | 15 | 1 | 15 | 2057.11431 | 2057.11399 | 0.00031 | 0.25 |
| 16 | 1 | 16 | 15 | 2 | 13 | 2057.11431 | 2057.11400 | 0.00030 | 0.25 |
| 8 | 4 | 5 | 7 | 3 | 4 | 2057.13517 | 2057.13538 | -0.00020 | 1.00 |
| 6 | 5 | 2 | 5 | 4 | 1 | 2057.14135 | 2057.14136 | -0.00000 | 0.25 |
| 6 | 5 | 1 | 5 | 4 | 2 | 2057.14135 | 2057.14150 | -0.00015 | 0.25 |
| 8 | 3 | 5 | 7 | 2 | 6 | 2057.15555 | 2057.15565 | -0.00010 | 1.00 |
| 9 | 4 | 6 | 8 | 3 | 5 | 2057.18222 | 2057.18201 | 0.00021 | 0.25 |
| 7 | 5 | 2 | 6 | 4 | 3 | 2057.21133 | 2057.21189 | -0.00056 | 0.25 |
| 7 | 5 | 3 | 6 | 4 | 2 | 2057.21133 | 2057.21118 | 0.00015 | 0.25 |
| 6 | 6 | 0 | 5 | 5 | 1 | 2057.26372 | 2057.26347 | 0.00025 | 0.25 |
| 6 | 6 | 1 | 5 | 5 | 0 | 2057.26372 | 2057.26346 | 0.00025 | 0.25 |
| 8 | 5 | 4 | 7 | 4 | 3 | 2057.27946 | 2057.27974 | -0.00027 | 1.00 |
| 8 | 5 | 3 | 7 | 4 | 4 | 2057.28250 | 2057.28233 | 0.00016 | 1.00 |
| 7 | 6 | 2 | 6 | 5 | 1 | 2057.33372 | 2057.33386 | -0.00014 | 0.25 |
| 7 | 6 | 1 | 6 | 5 | 2 | 2057.33372 | 2057.33388 | -0.00015 | 0.25 |

| 9 | 5 | 5 | 8 | 4 | 4 | 2057.34601 | 2057.34578 | 0.00023 | 0.25 |
|---|---|---|---|---|---|---|---|---|---|
| 9 | 5 | 4 | 8 | 4 | 5 | 2057.35372 | 2057.35342 | 0.00029 | 0.25 |
| 8 | 6 | 2 | 7 | 5 | 3 | 2057.40446 | 2057.40406 | 0.00039 | 0.25 |
| 8 | 6 | 3 | 7 | 5 | 2 | 2057.40446 | 2057.40399 | 0.00047 | 0.25 |
| 13 | 6 | 8 | 12 | 5 | 7 | 2057.73328 | 2057.73461 | -0.00133 | 0.00 |
| 10 | 7 | 3 | 11 | 8 | 4 | 2054.46520 | 2054.46885 | -0.00365 | 0.00 |
| 10 | 7 | 4 | 11 | 8 | 3 | 2054.46520 | 2054.46884 | -0.00364 | 0.00 |
| 11 | 7 | 4 | 12 | 8 | 5 | 2054.39494 | 2054.39724 | -0.00230 | 0.00 |
| 11 | 7 | 5 | 12 | 8 | 4 | 2054.39494 | 2054.39722 | -0.00228 | 0.00 |
| 12 | 7 | 5 | 13 | 8 | 6 | 2054.32388 | 2054.32583 | -0.00195 | 0.00 |
| 12 | 7 | 6 | 13 | 8 | 5 | 2054.32388 | 2054.32578 | -0.00190 | 0.00 |
| 12 | 7 | 5 | 11 | 6 | 6 | 2057.80230 | 2057.80476 | -0.00245 | 0.00 |
| 12 | 7 | 6 | 11 | 6 | 5 | 2057.80230 | 2057.80429 | -0.00199 | 0.00 |
| 13 | 7 | 6 | 14 | 8 | 7 | 2054.25311 | 2054.25477 | -0.00166 | 0.00 |
| 13 | 7 | 7 | 14 | 8 | 6 | 2054.25311 | 2054.25459 | -0.00148 | 0.00 |
| 9 | 8 | 1 | 8 | 7 | 2 | 2057.71409 | 2057.71834 | -0.00424 | 0.00 |
| 9 | 8 | 2 | 8 | 7 | 1 | 2057.71409 | 2057.71834 | -0.00424 | 0.00 |
| 9 | 8 | 1 | 10 | 9 | 2 | 2054.41467 | 2054.41912 | -0.00445 | 0.00 |
| 9 | 8 | 2 | 10 | 9 | 1 | 2054.41467 | 2054.41912 | -0.00445 | 0.00 |
| 10 | 8 | 2 | 9 | 7 | 3 | 2057.78517 | 2057.78840 | -0.00323 | 0.00 |
| 10 | 8 | 3 | 9 | 7 | 2 | 2057.78517 | 2057.78840 | -0.00323 | 0.00 |
| 10 | 8 | 2 | 11 | 9 | 3 | 2054.34389 | 2054.34725 | -0.00336 | 0.00 |
| 10 | 8 | 3 | 11 | 9 | 2 | 2054.34389 | 2054.34725 | -0.00336 | 0.00 |
| 11 | 8 | 3 | 10 | 7 | 4 | 2057.85517 | 2057.85823 | -0.00305 | 0.00 |
| 11 | 8 | 4 | 10 | 7 | 3 | 2057.85517 | 2057.85823 | -0.00305 | 0.00 |
| 11 | 8 | 3 | 12 | 9 | 4 | 2054.27241 | 2054.27539 | -0.00298 | 0.00 |
| 11 | 8 | 4 | 12 | 9 | 3 | 2054.27241 | 2054.27539 | -0.00298 | 0.00 |
| 12 | 8 | 4 | 11 | 7 | 5 | 2057.92511 | 2057.92777 | -0.00265 | 0.00 |
| 12 | 8 | 5 | 11 | 7 | 4 | 2057.92511 | 2057.92776 | -0.00264 | 0.00 |
| 12 | 8 | 4 | 13 | 9 | 5 | 2054.20064 | 2054.20360 | -0.00296 | 0.00 |
| 12 | 8 | 5 | 13 | 9 | 4 | 2054.20064 | 2054.20359 | -0.00295 | 0.00 |
| 13 | 8 | 5 | 12 | 7 | 6 | 2057.99424 | 2057.99694 | -0.00270 | 0.00 |
| 13 | 8 | 6 | 12 | 7 | 5 | 2057.99424 | 2057.99689 | -0.00264 | 0.00 |
| 9 | 9 | 0 | 8 | 8 | 1 | 2057.83664 | 2057.84024 | -0.00359 | 0.00 |
| 9 | 9 | 1 | 8 | 8 | 0 | 2057.83664 | 2057.84024 | -0.00359 | 0.00 |
| 9 | 9 | 0 | 10 | 10 | 1 | 2054.29437 | 2054.29783 | -0.00346 | 0.00 |
| 9 | 9 | 1 | 10 | 10 | 0 | 2054.29437 | 2054.29783 | -0.00346 | 0.00 |
| 10 | 9 | 1 | 9 | 8 | 2 | 2057.90713 | 2057.91037 | -0.00323 | 0.00 |
| 10 | 9 | 2 | 9 | 8 | 1 | 2057.90713 | 2057.91037 | -0.00323 | 0.00 |
| 10 | 9 | 1 | 11 | 10 | 2 | 2054.22240 | 2054.22592 | -0.00352 | 0.00 |
| 10 | 9 | 2 | 11 | 10 | 1 | 2054.22240 | 2054.22592 | -0.00352 | 0.00 |
| 11 | 9 | 2 | 10 | 8 | 3 | 2057.97724 | 2057.98034 | -0.00310 | 0.00 |
| 11 | 9 | 3 | 10 | 8 | 2 | 2057.97724 | 2057.98034 | -0.00310 | 0.00 |
| 13 | 9 | 4 | 14 | 10 | 5 | 2054.00690 | 2054.01007 | -0.00317 | 0.00 |
| 13 | 9 | 5 | 14 | 10 | 4 | 2054.00690 | 2054.01007 | -0.00317 | 0.00 |
| 11 | 10 | 1 | 10 | 9 | 2 | 2058.09945 | 2058.10225 | -0.00279 | 0.00 |
| 11 | 10 | 2 | 10 | 9 | 1 | 2058.09945 | 2058.10225 | -0.00279 | 0.00 |
| 11 | 10 | 1 | 12 | 11 | 2 | 2054.03009 | 2054.03277 | -0.00268 | 0.00 |
| 11 | 10 | 2 | 12 | 11 | 1 | 2054.03009 | 2054.03277 | -0.00268 | 0.00 |
| 12 | 10 | 2 | 11 | 9 | 3 | 2058.16962 | 2058.17213 | -0.00250 | 0.00 |
| 12 | 10 | 3 | 11 | 9 | 2 | 2058.16962 | 2058.17213 | -0.00250 | 0.00 |
| 13 | 10 | 3 | 12 | 9 | 4 | 2058.23988 | 2058.24184 | -0.00196 | 0.00 |
| 13 | 10 | 4 | 12 | 9 | 3 | 2058.23988 | 2058.24184 | -0.00196 | 0.00 |
| 14 | 10 | 4 | 13 | 9 | 5 | 2058.30437 | 2058.31134 | -0.00697 | 0.00 |
| 14 | 10 | 5 | 13 | 9 | 4 | 2058.30437 | 2058.31134 | -0.00697 | 0.00 |
| 11 | 11 | 0 | 10 | 10 | 1 | 2058.22057 | 2058.22406 | -0.00348 | 0.00 |
| 11 | 11 | 1 | 10 | 10 | 0 | 2058.22057 | 2058.22406 | -0.00348 | 0.00 |
| 12 | 11 | 1 | 11 | 10 | 2 | 2058.29059 | 2058.29397 | -0.00337 | 0.00 |
| 12 | 11 | 2 | 11 | 10 | 1 | 2058.29059 | 2058.29397 | -0.00337 | 0.00 |

****************************************************************************

Table A-2. Observed and calculated transitions for OCS-DCCCCD in the region of the $\nu_1$ fundamental band of OCS (units of cm$^{-1}$).

```
************************************************************************
J'  Ka' Kc'  J"  Ka" Kc"   Observed    Calculated  Obs-Calc   Weight
************************************************************************
11   5   6   12   6   7   2054.55716  2054.55717  -0.00001    1.00
 9   6   3   10   7   4   2054.57229  2054.57229  -0.00000    0.25
 9   6   4   10   7   3   2054.57229  2054.57224   0.00005    0.25
 7   7   0    8   8   1   2054.59903  2054.59874   0.00028    0.25
 7   7   1    8   8   0   2054.59903  2054.59874   0.00028    0.25
10   5   6   11   6   5   2054.61494  2054.61503  -0.00009    1.00
 8   6   2    9   7   3   2054.64101  2054.64100   0.00001    0.25
 8   6   3    9   7   2   2054.64101  2054.64099   0.00002    0.25
 7   6   1    8   7   2   2054.70992  2054.70987   0.00004    0.25
 7   6   2    8   7   1   2054.70992  2054.70987   0.00005    0.25
 8   5   3    9   6   4   2054.75274  2054.75293  -0.00018    0.25
 8   5   4    9   6   3   2054.75274  2054.75247   0.00027    0.25
10   4   6   11   5   7   2054.76227  2054.76242  -0.00014    1.00
 6   6   0    7   7   1   2054.77895  2054.77881   0.00013    0.25
 6   6   1    7   7   0   2054.77895  2054.77881   0.00013    0.25
 7   5   2    8   6   3   2054.82113  2054.82124  -0.00011    0.25
 7   5   3    8   6   2   2054.82113  2054.82114  -0.00000    0.25
 9   4   6   10   5   5   2054.78812  2054.78808   0.00004    1.00
 6   5   1    7   6   2   2054.89003  2054.88996   0.00007    0.25
 6   5   2    7   6   1   2054.89003  2054.88994   0.00009    0.25
 7   4   4    8   5   3   2054.93185  2054.93173   0.00012    1.00
 5   5   0    6   6   1   2054.95901  2054.95883   0.00018    0.25
 5   5   1    6   6   0   2054.95901  2054.95883   0.00018    0.25
 5   4   1    6   5   2   2055.07020  2055.07010   0.00009    0.25
 5   4   2    6   5   1   2055.07020  2055.06992   0.00027    0.25
```

| | | | | | | | | |
|---|---|---|---|---|---|---|---|---|
| 12 | 4 | 8 | 13 | 3 | 11 | 2055.09382 | 2055.09400 | -0.00018 | 1.00 |
| 6 | 3 | 3 | 7 | 4 | 4 | 2055.12487 | 2055.12454 | 0.00032 | 1.00 |
| 4 | 4 | 0 | 5 | 5 | 1 | 2055.13889 | 2055.13882 | 0.00006 | 0.25 |
| 4 | 4 | 1 | 5 | 5 | 0 | 2055.13889 | 2055.13880 | 0.00009 | 0.25 |
| 5 | 3 | 3 | 6 | 4 | 2 | 2055.17885 | 2055.17868 | 0.00016 | 0.25 |
| 5 | 3 | 2 | 6 | 4 | 3 | 2055.18534 | 2055.18533 | 0.00000 | 0.25 |
| 9 | 2 | 7 | 10 | 3 | 8 | 2055.19955 | 2055.19957 | -0.00002 | 1.00 |
| 10 | 1 | 9 | 11 | 2 | 10 | 2055.28005 | 2055.27936 | 0.00068 | 1.00 |
| 11 | 0 | 11 | 12 | 1 | 12 | 2055.28591 | 2055.28556 | 0.00034 | 0.25 |
| 11 | 1 | 11 | 12 | 0 | 12 | 2055.28591 | 2055.28570 | 0.00020 | 0.25 |
| 3 | 3 | 0 | 4 | 4 | 1 | 2055.31860 | 2055.31887 | -0.00027 | 0.25 |
| 3 | 3 | 1 | 4 | 4 | 0 | 2055.31860 | 2055.31863 | -0.00003 | 0.25 |
| 10 | 0 | 10 | 11 | 1 | 11 | 2055.34159 | 2055.34140 | 0.00018 | 0.25 |
| 10 | 1 | 10 | 11 | 0 | 11 | 2055.34159 | 2055.34173 | -0.00014 | 0.25 |
| 4 | 2 | 3 | 5 | 3 | 2 | 2055.34514 | 2055.34527 | -0.00013 | 1.00 |
| 9 | 0 | 9 | 10 | 1 | 10 | 2055.39754 | 2055.39709 | 0.00044 | 0.25 |
| 9 | 1 | 9 | 10 | 0 | 10 | 2055.39754 | 2055.39782 | -0.00028 | 0.25 |
| 8 | 2 | 7 | 9 | 1 | 8 | 2055.41435 | 2055.41467 | -0.00032 | 1.00 |
| 3 | 2 | 2 | 4 | 3 | 1 | 2055.42416 | 2055.42453 | -0.00037 | 1.00 |
| 3 | 2 | 1 | 4 | 3 | 2 | 2055.43607 | 2055.43604 | 0.00002 | 1.00 |
| 6 | 1 | 5 | 7 | 2 | 6 | 2055.45954 | 2055.45964 | -0.00010 | 1.00 |
| 7 | 0 | 7 | 8 | 1 | 8 | 2055.50741 | 2055.50738 | 0.00002 | 0.25 |
| 7 | 1 | 7 | 8 | 0 | 8 | 2055.51002 | 2055.51079 | -0.00077 | 0.25 |
| 6 | 0 | 6 | 7 | 1 | 7 | 2055.56121 | 2055.56129 | -0.00008 | 0.25 |
| 6 | 2 | 5 | 7 | 1 | 6 | 2055.56121 | 2055.56063 | 0.00057 | 0.25 |
| 1 | 1 | 0 | 1 | 0 | 1 | 2056.04460 | 2056.04444 | 0.00016 | 1.00 |
| 2 | 1 | 1 | 2 | 0 | 2 | 2056.05888 | 2056.05888 | -0.00000 | 1.00 |
| 2 | 0 | 2 | 1 | 1 | 1 | 2056.06663 | 2056.06658 | 0.00004 | 1.00 |
| 1 | 1 | 1 | 0 | 0 | 0 | 2056.09997 | 2056.09997 | -0.00000 | 1.00 |
| 4 | 1 | 3 | 5 | 2 | 4 | 2055.53668 | 2055.53666 | 0.00001 | 1.00 |
| 11 | 3 | 8 | 11 | 2 | 9 | 2055.53325 | 2055.53289 | 0.00035 | 1.00 |
| 6 | 1 | 6 | 7 | 0 | 7 | 2055.56817 | 2055.56836 | -0.00019 | 1.00 |

| | | | | | | | | |
|---|---|---|---|---|---|---|---|---|
| 9 | 3 | 7 | 9 | 4 | 6 | 2055.57106 | 2055.57128 | -0.00022 | 1.00 |
| 4 | 2 | 2 | 4 | 1 | 3 | 2056.11979 | 2056.11961 | 0.00018 | 1.00 |
| 3 | 0 | 3 | 2 | 1 | 2 | 2056.13946 | 2056.13899 | 0.00046 | 1.00 |
| 2 | 2 | 0 | 2 | 1 | 1 | 2056.13315 | 2056.13312 | 0.00003 | 1.00 |
| 6 | 2 | 4 | 6 | 1 | 5 | 2056.14468 | 2056.14491 | -0.00022 | 1.00 |
| 2 | 1 | 2 | 1 | 0 | 1 | 2056.15536 | 2056.15540 | -0.00004 | 1.00 |
| 2 | 2 | 1 | 2 | 1 | 2 | 2056.16830 | 2056.16850 | -0.00020 | 1.00 |
| 5 | 1 | 4 | 5 | 0 | 5 | 2056.17220 | 2056.17223 | -0.00003 | 1.00 |
| 7 | 2 | 5 | 7 | 1 | 6 | 2056.17954 | 2056.17930 | 0.00024 | 1.00 |
| 8 | 3 | 5 | 8 | 2 | 6 | 2056.18999 | 2056.19003 | -0.00003 | 1.00 |
| 7 | 3 | 4 | 7 | 2 | 5 | 2056.19231 | 2056.19265 | -0.00034 | 1.00 |
| 5 | 3 | 2 | 5 | 4 | 1 | 2055.59985 | 2055.59995 | -0.00010 | 1.00 |
| 5 | 0 | 5 | 6 | 1 | 6 | 2055.61391 | 2055.61357 | 0.00033 | 1.00 |
| 5 | 2 | 4 | 6 | 1 | 5 | 2055.64359 | 2055.64386 | -0.00027 | 1.00 |
| 4 | 0 | 4 | 5 | 1 | 5 | 2055.66344 | 2055.66358 | -0.00014 | 1.00 |
| 7 | 0 | 7 | 7 | 1 | 6 | 2055.66677 | 2055.66698 | -0.00021 | 1.00 |
| 6 | 1 | 6 | 6 | 2 | 5 | 2055.67429 | 2055.67481 | -0.00052 | 1.00 |
| 5 | 1 | 5 | 5 | 2 | 4 | 2055.71450 | 2055.71495 | -0.00045 | 1.00 |
| 4 | 2 | 2 | 4 | 3 | 1 | 2055.72270 | 2055.72267 | 0.00002 | 1.00 |
| 5 | 2 | 3 | 5 | 3 | 2 | 2055.73871 | 2055.73906 | -0.00035 | 1.00 |
| 4 | 1 | 4 | 4 | 2 | 3 | 2055.74874 | 2055.74885 | -0.00011 | 1.00 |
| 5 | 0 | 5 | 5 | 1 | 4 | 2055.79150 | 2055.79111 | 0.00038 | 1.00 |
| 2 | 1 | 2 | 2 | 2 | 1 | 2055.79590 | 2055.79598 | -0.00008 | 1.00 |
| 1 | 0 | 1 | 2 | 1 | 2 | 2055.80963 | 2055.80922 | 0.00040 | 1.00 |
| 6 | 1 | 5 | 6 | 2 | 4 | 2055.81790 | 2055.81758 | 0.00032 | 1.00 |
| 3 | 2 | 2 | 4 | 1 | 3 | 2055.82091 | 2055.82148 | -0.00056 | 1.00 |
| 2 | 1 | 2 | 3 | 0 | 3 | 2055.82577 | 2055.82540 | 0.00037 | 1.00 |
| 2 | 1 | 1 | 2 | 2 | 0 | 2055.83157 | 2055.83132 | 0.00025 | 1.00 |
| 0 | 0 | 0 | 1 | 1 | 1 | 2055.86479 | 2055.86479 | -0.00000 | 1.00 |
| 4 | 4 | 1 | 3 | 3 | 0 | 2056.64492 | 2056.64482 | 0.00009 | 0.25 |
| 4 | 4 | 0 | 3 | 3 | 1 | 2056.64492 | 2056.64506 | -0.00014 | 0.25 |
| 7 | 4 | 3 | 6 | 3 | 4 | 2056.85729 | 2056.85681 | 0.00047 | 1.00 |
| 6 | 4 | 3 | 5 | 3 | 2 | 2056.77731 | 2056.77727 | 0.00003 | 1.00 |
| 9 | 6 | 3 | 8 | 5 | 4 | 2057.20685 | 2057.20707 | -0.00022 | 0.25 |
| 9 | 6 | 4 | 8 | 5 | 3 | 2057.20685 | 2057.20661 | 0.00023 | 0.25 |
| 7 | 7 | 0 | 6 | 6 | 1 | 2057.18208 | 2057.18218 | -0.00010 | 0.25 |
| 7 | 7 | 1 | 6 | 6 | 0 | 2057.18208 | 2057.18218 | -0.00010 | 0.25 |
| 10 | 5 | 6 | 9 | 4 | 5 | 2057.14631 | 2057.14616 | 0.00014 | 0.25 |
| 10 | 4 | 6 | 9 | 3 | 7 | 2057.12740 | 2057.12777 | -0.00037 | 0.25 |
| 9 | 5 | 4 | 8 | 4 | 5 | 2057.09857 | 2057.09859 | -0.00002 | 1.00 |
| 7 | 6 | 1 | 6 | 5 | 2 | 2057.07136 | 2057.07142 | -0.00006 | 0.25 |
| 7 | 6 | 2 | 6 | 5 | 1 | 2057.07136 | 2057.07140 | -0.00004 | 0.25 |
| 10 | 8 | 2 | 9 | 7 | 3 | 2057.49658 | 2057.49662 | -0.00004 | 0.25 |
| 10 | 8 | 3 | 9 | 7 | 2 | 2057.49658 | 2057.49662 | -0.00004 | 0.25 |
| 9 | 8 | 1 | 8 | 7 | 2 | 2057.42877 | 2057.42883 | -0.00006 | 0.25 |
| 9 | 8 | 2 | 8 | 7 | 1 | 2057.42877 | 2057.42883 | -0.00006 | 0.25 |
| 10 | 7 | 3 | 9 | 6 | 4 | 2057.38581 | 2057.38574 | 0.00006 | 0.25 |
| 10 | 7 | 4 | 9 | 6 | 3 | 2057.38581 | 2057.38568 | 0.00012 | 0.25 |
| 8 | 8 | 0 | 7 | 7 | 1 | 2057.36076 | 2057.36088 | -0.00012 | 0.25 |
| 8 | 8 | 1 | 7 | 7 | 0 | 2057.36076 | 2057.36088 | -0.00012 | 0.25 |
| 11 | 6 | 5 | 10 | 5 | 6 | 2057.34179 | 2057.34220 | -0.00041 | 1.00 |
| 9 | 7 | 2 | 8 | 6 | 3 | 2057.31803 | 2057.31811 | -0.00008 | 0.25 |
| 9 | 7 | 3 | 8 | 6 | 2 | 2057.31803 | 2057.31810 | -0.00007 | 0.25 |
| 13 | 5 | 9 | 12 | 2 | 10 | 2057.25971 | 2057.25985 | -0.00014 | 1.00 |
| 11 | 5 | 7 | 10 | 4 | 6 | 2057.19538 | 2057.19515 | 0.00022 | 1.00 |
| 8 | 5 | 3 | 7 | 4 | 4 | 2057.02941 | 2057.02911 | 0.00029 | 1.00 |
| 14 | 4 | 11 | 13 | 1 | 12 | 2057.04275 | 2057.04245 | 0.00029 | 1.00 |
| 8 | 6 | 2 | 7 | 5 | 3 | 2057.13926 | 2057.13937 | -0.00011 | 0.25 |
| 8 | 6 | 3 | 7 | 5 | 2 | 2057.13926 | 2057.13926 | -0.00000 | 0.25 |
| 11 | 4 | 8 | 10 | 3 | 7 | 2056.98823 | 2056.98815 | 0.00007 | 1.00 |
| 6 | 6 | 0 | 5 | 5 | 1 | 2057.00303 | 2057.00327 | -0.00024 | 0.25 |
| 6 | 6 | 1 | 5 | 5 | 0 | 2057.00303 | 2057.00327 | -0.00024 | 0.25 |
| 7 | 5 | 3 | 6 | 4 | 2 | 2056.96040 | 2056.95975 | 0.00064 | 0.25 |
| 7 | 5 | 2 | 6 | 4 | 3 | 2056.96040 | 2056.96068 | -0.00028 | 0.25 |
| 8 | 4 | 4 | 7 | 3 | 5 | 2056.93556 | 2056.93548 | 0.00007 | 0.25 |
| 8 | 3 | 5 | 7 | 2 | 6 | 2056.93556 | 2056.93502 | 0.00053 | 0.25 |
| 9 | 4 | 6 | 8 | 3 | 5 | 2056.93181 | 2056.93174 | 0.00006 | 1.00 |
| 6 | 5 | 2 | 5 | 4 | 1 | 2056.89257 | 2056.89228 | 0.00028 | 0.25 |
| 6 | 5 | 1 | 5 | 4 | 2 | 2056.89257 | 2056.89247 | 0.00009 | 0.25 |
| 8 | 4 | 5 | 7 | 3 | 4 | 2056.88966 | 2056.88950 | 0.00015 | 1.00 |
| 10 | 3 | 8 | 9 | 2 | 7 | 2056.76016 | 2056.75982 | 0.00033 | 1.00 |
| 13 | 1 | 13 | 12 | 0 | 12 | 2056.72816 | 2056.72838 | -0.00022 | 0.25 |
| 13 | 0 | 13 | 12 | 1 | 12 | 2056.72816 | 2056.72832 | -0.00016 | 0.25 |
| 11 | 2 | 10 | 10 | 1 | 9 | 2056.67985 | 2056.67972 | 0.00012 | 1.00 |
| 11 | 3 | 8 | 10 | 2 | 9 | 2056.66947 | 2056.66913 | 0.00033 | 1.00 |
| 11 | 1 | 11 | 10 | 0 | 10 | 2056.61810 | 2056.61849 | -0.00039 | 0.25 |
| 11 | 0 | 11 | 10 | 1 | 10 | 2056.61810 | 2056.61817 | -0.00007 | 0.25 |
| 10 | 1 | 9 | 9 | 2 | 8 | 2056.60957 | 2056.60950 | 0.00006 | 1.00 |
| 10 | 1 | 10 | 9 | 0 | 9 | 2056.56307 | 2056.56363 | -0.00056 | 0.25 |
| 10 | 0 | 10 | 9 | 1 | 9 | 2056.56307 | 2056.56291 | 0.00015 | 0.25 |
| 4 | 3 | 2 | 3 | 2 | 1 | 2056.52788 | 2056.52779 | 0.00008 | 1.00 |
| 8 | 1 | 7 | 7 | 2 | 6 | 2056.47694 | 2056.47739 | -0.00045 | 1.00 |
| 9 | 1 | 8 | 8 | 2 | 7 | 2056.54552 | 2056.54619 | -0.00067 | 1.00 |
| 8 | 1 | 8 | 7 | 0 | 7 | 2056.45488 | 2056.45477 | 0.00010 | 1.00 |
| 8 | 0 | 8 | 7 | 1 | 7 | 2056.45132 | 2056.45136 | -0.00004 | 1.00 |
| 9 | 4 | 6 | 9 | 3 | 7 | 2056.38848 | 2056.38849 | -0.00001 | 1.00 |
| 7 | 1 | 7 | 6 | 0 | 6 | 2056.40113 | 2056.40146 | -0.00033 | 0.25 |
| 7 | 1 | 6 | 6 | 2 | 5 | 2056.40113 | 2056.40167 | -0.00054 | 0.25 |
| 6 | 4 | 2 | 6 | 3 | 3 | 2056.35278 | 2056.35309 | -0.00031 | 1.00 |
| 6 | 1 | 6 | 5 | 0 | 5 | 2056.34946 | 2056.34970 | -0.00024 | 1.00 |
| 5 | 0 | 5 | 4 | 1 | 4 | 2056.27446 | 2056.27422 | 0.00024 | 1.00 |
| 3 | 3 | 1 | 3 | 2 | 2 | 2056.26223 | 2056.26246 | -0.00023 | 1.00 |
| 5 | 2 | 4 | 5 | 1 | 5 | 2056.24839 | 2056.24843 | -0.00004 | 1.00 |
| 8 | 2 | 6 | 8 | 1 | 7 | 2056.22874 | 2056.22859 | 0.00015 | 1.00 |
| 4 | 2 | 3 | 4 | 1 | 4 | 2056.21514 | 2056.21498 | 0.00015 | 1.00 |
| 4 | 0 | 4 | 3 | 1 | 3 | 2056.20858 | 2056.20875 | -0.00016 | 1.00 |

| J' | Ka' | Kc' | J" | Ka" | Kc" | Observed | Calculated | Obs-Calc | Weight |
|---|---|---|---|---|---|---|---|---|---|
| 9 | 3 | 6 | 9 | 2 | 7 | 2056.20198 | 2056.20220 | -0.00021 | 1.00 |
| 1 | 0 | 1 | 1 | 1 | 0 | 2055.92063 | 2055.92026 | 0.00036 | 1.00 |
| 2 | 0 | 2 | 2 | 1 | 1 | 2055.90586 | 2055.90561 | 0.00024 | 1.00 |
| 3 | 0 | 3 | 3 | 1 | 2 | 2055.88088 | 2055.88024 | 0.00064 | 1.00 |
| 7 | 1 | 6 | 7 | 2 | 5 | 2055.78206 | 2055.78248 | -0.00042 | 1.00 |
| 3 | 1 | 3 | 3 | 2 | 2 | 2055.77564 | 2055.77593 | -0.00029 | 1.00 |

****************************************************************************

Table A-3. Observed and calculated transitions for N$_2$O-HCCCH in the region of the $\nu_1$ fundamental band of N$_2$O (in units of cm$^{-1}$).

****************************************************************************

| J' | Ka' | Kc' | J" | Ka" | Kc" | Observed | Calculated | Obs-Calc | Weight |
|---|---|---|---|---|---|---|---|---|---|
| 8 | 7 | 1 | 9 | 8 | 2 | 2223.65138 | 2223.65179 | -0.00041 | 0.25 |
| 8 | 7 | 2 | 9 | 8 | 1 | 2223.65138 | 2223.65179 | -0.00041 | 0.25 |
| 7 | 7 | 1 | 8 | 8 | 0 | 2223.73462 | 2223.73466 | -0.00004 | 0.25 |
| 7 | 7 | 0 | 8 | 8 | 1 | 2223.73462 | 2223.73466 | -0.00004 | 0.25 |
| 10 | 5 | 6 | 11 | 6 | 5 | 2223.93132 | 2223.93177 | -0.00045 | 0.25 |
| 10 | 5 | 5 | 11 | 6 | 6 | 2223.93132 | 2223.93197 | -0.00065 | 0.25 |
| 7 | 6 | 2 | 8 | 7 | 1 | 2223.95739 | 2223.95735 | 0.00004 | 0.25 |
| 7 | 6 | 1 | 8 | 7 | 2 | 2223.95739 | 2223.95735 | 0.00004 | 0.25 |
| 9 | 5 | 4 | 10 | 6 | 5 | 2224.01392 | 2224.01420 | -0.00028 | 0.25 |
| 9 | 5 | 5 | 10 | 6 | 4 | 2224.01392 | 2224.01413 | -0.00021 | 0.25 |
| 8 | 5 | 4 | 9 | 6 | 3 | 2224.09656 | 2224.09675 | -0.00018 | 0.25 |
| 8 | 5 | 3 | 9 | 6 | 4 | 2224.09656 | 2224.09677 | -0.00020 | 0.25 |
| 7 | 5 | 2 | 8 | 6 | 3 | 2224.17925 | 2224.17956 | -0.00030 | 0.25 |
| 7 | 5 | 3 | 8 | 6 | 2 | 2224.17925 | 2224.17955 | -0.00030 | 0.25 |
| 6 | 5 | 1 | 7 | 6 | 2 | 2224.26267 | 2224.26248 | 0.00019 | 0.25 |
| 6 | 5 | 2 | 7 | 6 | 1 | 2224.26267 | 2224.26248 | 0.00019 | 0.25 |

| | | | | | | | | |
|---|---|---|---|---|---|---|---|---|
| 8 | 4 | 4 | 9 | 5 | 5 | 2224.31880 | 2224.31958 | -0.00077 | 0.25 |
| 8 | 4 | 5 | 9 | 5 | 4 | 2224.31880 | 2224.31865 | 0.00014 | 0.25 |
| 5 | 5 | 0 | 6 | 6 | 1 | 2224.34583 | 2224.34548 | 0.00035 | 0.25 |
| 5 | 5 | 1 | 6 | 6 | 0 | 2224.34583 | 2224.34548 | 0.00035 | 0.25 |
| 6 | 4 | 2 | 7 | 5 | 3 | 2224.48427 | 2224.48417 | 0.00009 | 0.25 |
| 6 | 4 | 3 | 7 | 5 | 2 | 2224.48427 | 2224.48409 | 0.00018 | 0.25 |
| 8 | 3 | 6 | 9 | 4 | 5 | 2224.53313 | 2224.53329 | -0.00015 | 1.00 |
| 4 | 4 | 0 | 5 | 5 | 1 | 2224.65028 | 2224.65004 | 0.00024 | 0.25 |
| 4 | 4 | 1 | 5 | 5 | 0 | 2224.65028 | 2224.65004 | 0.00024 | 0.25 |
| 8 | 2 | 7 | 9 | 3 | 6 | 2224.67479 | 2224.67467 | 0.00011 | 1.00 |
| 6 | 3 | 4 | 7 | 4 | 3 | 2224.70375 | 2224.70388 | -0.00013 | 1.00 |
| 6 | 3 | 3 | 7 | 4 | 4 | 2224.70755 | 2224.70775 | -0.00019 | 1.00 |
| 5 | 3 | 2 | 6 | 4 | 3 | 2224.78805 | 2224.78875 | -0.00070 | 0.25 |
| 5 | 3 | 3 | 6 | 4 | 2 | 2224.78805 | 2224.78747 | 0.00058 | 0.25 |
| 9 | 2 | 7 | 10 | 3 | 8 | 2224.80968 | 2224.80981 | -0.00012 | 1.00 |
| 11 | 5 | 6 | 11 | 6 | 5 | 2224.85303 | 2224.85186 | 0.00116 | 1.00 |
| 4 | 3 | 1 | 5 | 4 | 2 | 2224.87090 | 2224.87101 | -0.00010 | 0.25 |
| 4 | 3 | 2 | 5 | 4 | 1 | 2224.87090 | 2224.87069 | 0.00021 | 0.25 |
| 6 | 2 | 5 | 7 | 3 | 4 | 2224.89668 | 2224.89693 | -0.00025 | 1.00 |
| 7 | 2 | 5 | 8 | 3 | 6 | 2224.90338 | 2224.90342 | -0.00004 | 1.00 |
| 3 | 3 | 1 | 4 | 4 | 0 | 2224.95399 | 2224.95380 | 0.00018 | 0.25 |
| 3 | 3 | 0 | 4 | 4 | 1 | 2224.95399 | 2224.95385 | 0.00014 | 0.25 |
| 6 | 2 | 4 | 7 | 3 | 5 | 2224.96127 | 2224.96114 | 0.00013 | 1.00 |
| 5 | 2 | 4 | 6 | 3 | 3 | 2224.99364 | 2224.99374 | -0.00010 | 1.00 |
| 13 | 0 | 13 | 14 | 1 | 14 | 2225.00922 | 2225.00769 | 0.00152 | 0.25 |
| 13 | 1 | 13 | 14 | 0 | 14 | 2225.00922 | 2225.00969 | -0.00047 | 0.25 |
| 4 | 2 | 2 | 5 | 3 | 3 | 2225.09870 | 2225.09873 | -0.00003 | 1.00 |
| 3 | 2 | 1 | 4 | 3 | 2 | 2225.17593 | 2225.17627 | -0.00034 | 1.00 |
| 10 | 0 | 10 | 11 | 1 | 11 | 2225.22035 | 2225.22048 | -0.00012 | 1.00 |
| 10 | 1 | 10 | 11 | 0 | 11 | 2225.22991 | 2225.22988 | 0.00002 | 1.00 |
| 2 | 2 | 0 | 3 | 3 | 1 | 2225.25661 | 2225.25718 | -0.00057 | 0.25 |
| 2 | 2 | 1 | 3 | 3 | 0 | 2225.25661 | 2225.25625 | 0.00035 | 0.25 |
| 6 | 1 | 5 | 7 | 2 | 6 | 2225.26118 | 2225.26150 | -0.00032 | 1.00 |
| 9 | 1 | 9 | 10 | 0 | 10 | 2225.30499 | 2225.30486 | 0.00013 | 1.00 |
| 5 | 1 | 4 | 6 | 2 | 5 | 2225.31179 | 2225.31202 | -0.00023 | 1.00 |
| 8 | 0 | 8 | 9 | 1 | 9 | 2225.35692 | 2225.35704 | -0.00011 | 1.00 |
| 8 | 1 | 8 | 9 | 1 | 9 | 2225.37214 | 2225.37180 | 0.00033 | 1.00 |
| 7 | 3 | 5 | 8 | 3 | 6 | 2225.37679 | 2225.37723 | -0.00043 | 1.00 |
| 8 | 1 | 8 | 9 | 0 | 9 | 2225.38138 | 2225.38138 | 0.00000 | 1.00 |
| 7 | 0 | 7 | 8 | 1 | 8 | 2225.42222 | 2225.42241 | -0.00019 | 1.00 |
| 3 | 1 | 2 | 4 | 2 | 3 | 2225.42703 | 2225.42715 | -0.00012 | 1.00 |
| 7 | 1 | 7 | 8 | 0 | 8 | 2225.46038 | 2225.46009 | 0.00028 | 1.00 |
| 6 | 0 | 6 | 7 | 1 | 7 | 2225.48526 | 2225.48536 | -0.00010 | 1.00 |
| 6 | 2 | 5 | 6 | 3 | 4 | 2225.48900 | 2225.48944 | -0.00043 | 1.00 |
| 2 | 1 | 1 | 3 | 2 | 2 | 2225.49287 | 2225.49287 | 0.00000 | 1.00 |
| 4 | 2 | 2 | 4 | 3 | 1 | 2225.51453 | 2225.51461 | -0.00008 | 1.00 |
| 6 | 1 | 6 | 7 | 1 | 7 | 2225.51875 | 2225.51868 | 0.00006 | 1.00 |
| 5 | 0 | 5 | 6 | 0 | 6 | 2225.57996 | 2225.57991 | 0.00004 | 1.00 |
| 8 | 2 | 6 | 8 | 3 | 5 | 2225.58605 | 2225.58583 | 0.00021 | 1.00 |
| 5 | 1 | 5 | 6 | 1 | 6 | 2225.59292 | 2225.59290 | 0.00002 | 1.00 |
| 6 | 1 | 6 | 6 | 2 | 5 | 2225.59977 | 2225.59967 | 0.00009 | 1.00 |
| 4 | 0 | 4 | 5 | 1 | 5 | 2225.60561 | 2225.60546 | 0.00014 | 1.00 |
| 4 | 1 | 3 | 5 | 1 | 4 | 2225.61265 | 2225.61256 | 0.00009 | 1.00 |
| 4 | 2 | 2 | 5 | 2 | 3 | 2225.62060 | 2225.62047 | 0.00013 | 1.00 |
| 8 | 0 | 8 | 8 | 1 | 7 | 2225.64611 | 2225.64571 | 0.00039 | 1.00 |
| 4 | 0 | 4 | 5 | 0 | 5 | 2225.65296 | 2225.65291 | 0.00005 | 1.00 |
| 10 | 1 | 9 | 10 | 2 | 8 | 2225.68326 | 2225.68259 | 0.00067 | 0.25 |
| 10 | 1 | 9 | 10 | 2 | 8 | 2225.68326 | 2225.68259 | 0.00067 | 0.25 |
| 3 | 1 | 3 | 3 | 2 | 2 | 2225.68968 | 2225.68947 | 0.00020 | 1.00 |
| 3 | 1 | 2 | 4 | 1 | 3 | 2225.69874 | 2225.69856 | 0.00018 | 1.00 |
| 2 | 1 | 1 | 2 | 2 | 0 | 2225.74070 | 2225.74014 | 0.00055 | 1.00 |
| 3 | 1 | 3 | 4 | 1 | 4 | 2225.74337 | 2225.74322 | 0.00015 | 1.00 |
| 3 | 1 | 2 | 3 | 2 | 1 | 2225.75381 | 2225.75299 | 0.00082 | 0.25 |
| 3 | 1 | 2 | 3 | 2 | 1 | 2225.75381 | 2225.75299 | 0.00082 | 0.25 |

| | | | | | | | | |
|---|---|---|---|---|---|---|---|---|
| 8 | 1 | 7 | 8 | 2 | 6 | 2225.75635 | 2225.75599 | 0.00035 | 0.25 |
| 8 | 1 | 7 | 8 | 2 | 6 | 2225.75635 | 2225.75599 | 0.00035 | 0.25 |
| 4 | 1 | 3 | 4 | 2 | 2 | 2225.76614 | 2225.76583 | 0.00031 | 1.00 |
| 1 | 0 | 1 | 2 | 1 | 2 | 2225.79369 | 2225.79355 | 0.00014 | 1.00 |
| 2 | 1 | 2 | 3 | 1 | 3 | 2225.81918 | 2225.81933 | -0.00015 | 1.00 |
| 0 | 0 | 0 | 1 | 1 | 1 | 2225.86480 | 2225.86501 | -0.00021 | 1.00 |
| 4 | 0 | 4 | 4 | 1 | 3 | 2225.87445 | 2225.87490 | -0.00044 | 1.00 |
| 1 | 0 | 1 | 2 | 0 | 2 | 2225.88606 | 2225.88620 | -0.00013 | 1.00 |
| 3 | 0 | 3 | 3 | 1 | 2 | 2225.90388 | 2225.90424 | -0.00036 | 1.00 |
| 2 | 0 | 2 | 2 | 1 | 1 | 2225.92405 | 2225.92419 | -0.00014 | 1.00 |
| 1 | 0 | 1 | 1 | 1 | 0 | 2225.93642 | 2225.93642 | -0.00000 | 1.00 |
| 3 | 2 | 1 | 3 | 2 | 2 | 2226.05264 | 2226.05259 | 0.00004 | 0.25 |
| 7 | 3 | 4 | 7 | 3 | 5 | 2226.05264 | 2226.05364 | -0.00099 | 0.25 |
| 1 | 0 | 1 | 0 | 0 | 0 | 2226.13349 | 2226.13347 | 0.00001 | 1.00 |
| 1 | 1 | 0 | 1 | 0 | 1 | 2226.16425 | 2226.16450 | -0.00024 | 1.00 |
| 2 | 1 | 1 | 2 | 0 | 2 | 2226.17649 | 2226.17664 | -0.00015 | 1.00 |
| 3 | 1 | 2 | 3 | 0 | 3 | 2226.19634 | 2226.19643 | -0.00009 | 1.00 |
| 5 | 1 | 4 | 5 | 0 | 5 | 2226.26530 | 2226.26546 | -0.00015 | 1.00 |
| 3 | 1 | 3 | 2 | 1 | 2 | 2226.28063 | 2226.28060 | 0.00003 | 1.00 |
| 3 | 2 | 1 | 2 | 2 | 0 | 2226.29982 | 2226.29987 | -0.00005 | 1.00 |
| 2 | 1 | 2 | 1 | 0 | 1 | 2226.30713 | 2226.30724 | -0.00010 | 1.00 |
| 4 | 2 | 2 | 4 | 1 | 3 | 2226.33238 | 2226.33245 | -0.00006 | 1.00 |
| 8 | 2 | 6 | 8 | 1 | 7 | 2226.34160 | 2226.34159 | 0.00001 | 1.00 |
| 3 | 2 | 1 | 3 | 1 | 2 | 2226.34524 | 2226.34536 | -0.00011 | 1.00 |
| 2 | 2 | 1 | 2 | 1 | 2 | 2226.39118 | 2226.39123 | -0.00004 | 1.00 |
| 4 | 1 | 3 | 3 | 1 | 2 | 2226.40149 | 2226.40160 | -0.00010 | 1.00 |
| 3 | 2 | 2 | 3 | 1 | 3 | 2226.40890 | 2226.40861 | 0.00029 | 1.00 |
| 10 | 2 | 8 | 10 | 1 | 9 | 2226.41372 | 2226.41364 | 0.00007 | 1.00 |
| 4 | 1 | 4 | 3 | 0 | 3 | 2226.43526 | 2226.43504 | 0.00022 | 1.00 |
| 5 | 0 | 5 | 4 | 0 | 4 | 2226.44760 | 2226.44751 | 0.00009 | 1.00 |
| 8 | 1 | 7 | 8 | 0 | 8 | 2226.45220 | 2226.45202 | 0.00017 | 1.00 |
| 6 | 0 | 6 | 5 | 1 | 5 | 2226.47244 | 2226.47234 | 0.00010 | 1.00 |
| 5 | 2 | 3 | 4 | 2 | 2 | 2226.47626 | 2226.47602 | 0.00024 | 1.00 |
| 6 | 0 | 6 | 5 | 0 | 5 | 2226.51996 | 2226.51978 | 0.00017 | 1.00 |
| 2 | 2 | 1 | 1 | 1 | 0 | 2226.53431 | 2226.53410 | 0.00020 | 1.00 |
| 6 | 1 | 6 | 5 | 0 | 5 | 2226.55302 | 2226.55311 | -0.00008 | 1.00 |
| 7 | 0 | 7 | 6 | 1 | 6 | 2226.55645 | 2226.55654 | -0.00009 | 1.00 |
| 6 | 1 | 5 | 5 | 1 | 4 | 2226.57185 | 2226.57171 | 0.00013 | 1.00 |
| 5 | 3 | 3 | 5 | 2 | 4 | 2226.59617 | 2226.59571 | 0.00046 | 1.00 |
| 7 | 2 | 6 | 6 | 2 | 5 | 2226.61935 | 2226.61978 | -0.00042 | 1.00 |
| 8 | 0 | 8 | 7 | 1 | 7 | 2226.63751 | 2226.63718 | 0.00032 | 1.00 |
| 3 | 2 | 1 | 2 | 1 | 2 | 2226.64379 | 2226.64371 | 0.00007 | 1.00 |
| 4 | 2 | 3 | 3 | 1 | 2 | 2226.67106 | 2226.67104 | 0.00001 | 1.00 |
| 8 | 1 | 8 | 7 | 0 | 7 | 2226.67504 | 2226.67497 | 0.00006 | 1.00 |
| 9 | 1 | 9 | 8 | 0 | 8 | 2226.73920 | 2226.73922 | -0.00002 | 1.00 |
| 6 | 2 | 5 | 5 | 1 | 4 | 2226.78573 | 2226.78568 | 0.00004 | 1.00 |
| 10 | 0 | 10 | 9 | 1 | 9 | 2226.78977 | 2226.79008 | -0.00030 | 1.00 |
| 10 | 1 | 10 | 9 | 0 | 9 | 2226.80526 | 2226.80544 | -0.00017 | 1.00 |
| 11 | 0 | 11 | 10 | 1 | 10 | 2226.86323 | 2226.86370 | -0.00046 | 1.00 |
| 4 | 3 | 1 | 3 | 2 | 2 | 2226.92169 | 2226.92174 | -0.00005 | 1.00 |
| 5 | 3 | 3 | 4 | 2 | 2 | 2226.99444 | 2226.99454 | -0.00009 | 1.00 |
| 6 | 3 | 4 | 5 | 2 | 3 | 2227.06689 | 2227.06671 | 0.00018 | 1.00 |
| 6 | 3 | 3 | 5 | 2 | 4 | 2227.09924 | 2227.09923 | 0.00000 | 1.00 |
| 4 | 4 | 1 | 3 | 3 | 0 | 2227.13235 | 2227.13224 | 0.00010 | 0.25 |
| 4 | 4 | 0 | 3 | 3 | 1 | 2227.13235 | 2227.13229 | 0.00005 | 0.25 |
| 5 | 4 | 1 | 4 | 3 | 2 | 2227.21530 | 2227.21529 | 0.00001 | 0.25 |
| 5 | 4 | 2 | 4 | 3 | 1 | 2227.21530 | 2227.21497 | 0.00032 | 0.25 |
| 6 | 4 | 3 | 5 | 3 | 2 | 2227.29774 | 2227.29712 | 0.00061 | 0.25 |
| 6 | 4 | 2 | 5 | 3 | 3 | 2227.29774 | 2227.29839 | -0.00065 | 0.25 |
| 7 | 4 | 4 | 6 | 3 | 3 | 2227.37801 | 2227.37803 | -0.00002 | 1.00 |
| 7 | 4 | 3 | 6 | 3 | 4 | 2227.38185 | 2227.38184 | 0.00001 | 1.00 |
| 5 | 5 | 0 | 4 | 4 | 1 | 2227.42659 | 2227.42642 | 0.00016 | 0.25 |
| 5 | 5 | 1 | 4 | 4 | 0 | 2227.42659 | 2227.42642 | 0.00016 | 0.25 |
| 8 | 4 | 5 | 7 | 3 | 4 | 2227.45644 | 2227.45667 | -0.00022 | 1.00 |
| 6  | 5 | 1 | 5  | 4 | 2 | 2227.50945 | 2227.50938 |  0.00007 | 0.25 |
| 6  | 5 | 2 | 5  | 4 | 1 | 2227.50945 | 2227.50936 |  0.00009 | 0.25 |
| 9  | 4 | 6 | 8  | 3 | 5 | 2227.53119 | 2227.53156 | -0.00037 | 1.00 |
| 7  | 5 | 2 | 6  | 4 | 3 | 2227.59209 | 2227.59224 | -0.00015 | 0.25 |
| 7  | 5 | 3 | 6  | 4 | 2 | 2227.59209 | 2227.59216 | -0.00007 | 0.25 |
| 11 | 4 | 8 | 10 | 3 | 7 | 2227.66293 | 2227.66304 | -0.00010 | 1.00 |
| 8  | 5 | 4 | 7  | 4 | 3 | 2227.67511 | 2227.67468 |  0.00042 | 0.25 |
| 8  | 5 | 3 | 7  | 4 | 4 | 2227.67511 | 2227.67498 |  0.00012 | 0.25 |
| 9  | 5 | 4 | 8  | 4 | 5 | 2227.75734 | 2227.75757 | -0.00023 | 0.25 |
| 9  | 5 | 5 | 8  | 4 | 4 | 2227.75734 | 2227.75667 |  0.00066 | 0.25 |
| 7  | 6 | 2 | 6  | 5 | 1 | 2227.80253 | 2227.80224 |  0.00028 | 0.25 |
| 7  | 6 | 1 | 6  | 5 | 2 | 2227.80253 | 2227.80224 |  0.00028 | 0.25 |
| 8  | 6 | 3 | 7  | 5 | 2 | 2227.88545 | 2227.88521 |  0.00024 | 0.25 |
| 8  | 6 | 2 | 7  | 5 | 3 | 2227.88545 | 2227.88521 |  0.00023 | 0.25 |
| 11 | 5 | 7 | 10 | 4 | 6 | 2227.91733 | 2227.91733 | -0.00000 | 1.00 |
| 11 | 5 | 6 | 10 | 4 | 7 | 2227.92221 | 2227.92276 | -0.00054 | 1.00 |
| 9  | 6 | 4 | 8  | 5 | 3 | 2227.96819 | 2227.96805 |  0.00013 | 0.25 |
| 9  | 6 | 3 | 8  | 5 | 4 | 2227.96819 | 2227.96807 |  0.00011 | 0.25 |
| 7  | 7 | 0 | 6  | 6 | 1 | 2228.01059 | 2228.01067 | -0.00007 | 0.25 |
| 7  | 7 | 1 | 6  | 6 | 0 | 2228.01059 | 2228.01067 | -0.00007 | 0.25 |
| 10 | 6 | 5 | 9  | 5 | 4 | 2228.05081 | 2228.05071 |  0.00010 | 0.25 |
| 10 | 6 | 4 | 9  | 5 | 5 | 2228.05081 | 2228.05078 |  0.00003 | 0.25 |

****************************************************************************

Table A-4. Observed and calculated transitions for $N_2O$-DCCCCD in the region of the $v_1$ fundamental band of $N_2O$ (units of cm$^{-1}$).

***********************************************************************

| J' | Ka' | Kc' | J" | Ka" | Kc" | Observed | Calculated | Obs-Calc | Weight |
|----|-----|-----|----|-----|-----|----------|------------|----------|--------|
| 4 | 4 | 1 | 5 | 5 | 0 | 2224.78432 | 2224.78432 | -0.000001 | 0.25 |
| 4 | 4 | 0 | 5 | 5 | 1 | 2224.78432 | 2224.78432 | -0.000004 | 0.25 |
| 6 | 3 | 4 | 7 | 4 | 3 | 2224.82286 | 2224.82278 | 0.000086 | 1.00 |
| 6 | 3 | 3 | 7 | 4 | 4 | 2224.82767 | 2224.82750 | 0.000168 | 1.00 |
| 4 | 3 | 2 | 5 | 4 | 1 | 2224.98345 | 2224.98323 | 0.000226 | 0.25 |
| 4 | 3 | 1 | 5 | 4 | 2 | 2224.98345 | 2224.98362 | -0.000165 | 0.25 |
| 6 | 2 | 5 | 7 | 3 | 4 | 2224.99255 | 2224.99153 | 0.001019 | 1.00 |
| 7 | 2 | 5 | 8 | 3 | 6 | 2225.00849 | 2225.00895 | -0.000466 | 1.00 |
| 5 | 2 | 4 | 6 | 3 | 3 | 2225.08651 | 2225.08668 | -0.000174 | 1.00 |
| 4 | 2 | 3 | 5 | 3 | 2 | 2225.17501 | 2225.17514 | -0.000130 | 1.00 |
| 4 | 2 | 2 | 5 | 3 | 3 | 2225.19063 | 2225.19080 | -0.000170 | 1.00 |
| 8 | 1 | 7 | 9 | 2 | 8 | 2225.24995 | 2225.24986 | 0.000088 | 1.00 |
| 3 | 2 | 2 | 4 | 3 | 1 | 2225.25906 | 2225.25924 | -0.000176 | 1.00 |
| 9 | 1 | 8 | 10 | 1 | 9 | 2225.27476 | 2225.27445 | 0.000311 | 1.00 |
| 8 | 2 | 6 | 9 | 2 | 7 | 2225.31020 | 2225.31024 | -0.000041 | 1.00 |
| 8 | 3 | 5 | 9 | 3 | 6 | 2225.32742 | 2225.32718 | 0.000240 | 1.00 |
| 9 | 1 | 9 | 10 | 0 | 10 | 2225.35989 | 2225.35982 | 0.000076 | 1.00 |
| 7 | 0 | 7 | 8 | 1 | 8 | 2225.47724 | 2225.47739 | -0.000153 | 1.00 |
| 6 | 2 | 4 | 7 | 2 | 5 | 2225.48253 | 2225.48282 | -0.000291 | 1.00 |
| 7 | 0 | 7 | 8 | 0 | 8 | 2225.48845 | 2225.48860 | -0.000154 | 1.00 |
| 2 | 1 | 2 | 3 | 2 | 1 | 2225.51752 | 2225.51763 | -0.000109 | 1.00 |
| 8 | 2 | 7 | 8 | 3 | 6 | 2225.53152 | 2225.53176 | -0.000245 | 1.00 |
| 6 | 0 | 6 | 7 | 1 | 7 | 2225.53851 | 2225.53832 | 0.000190 | 1.00 |
| 4 | 2 | 3 | 4 | 3 | 2 | 2225.57613 | 2225.57614 | -0.000015 | 1.00 |
| 5 | 0 | 5 | 6 | 1 | 6 | 2225.59733 | 2225.59690 | 0.000436 | 1.00 |
| 5 | 2 | 3 | 5 | 3 | 2 | 2225.60158 | 2225.60148 | 0.000092 | 1.00 |
| 5 | 0 | 5 | 6 | 0 | 6 | 2225.62405 | 2225.62394 | 0.000107 | 1.00 |
| 5 | 1 | 5 | 6 | 1 | 6 | 2225.63543 | 2225.63554 | -0.000115 | 1.00 |
| 7 | 2 | 5 | 7 | 3 | 4 | 2225.63939 | 2225.63941 | -0.000019 | 1.00 |
| 4 | 2 | 2 | 5 | 2 | 3 | 2225.65825 | 2225.65833 | -0.000080 | 1.00 |
| 5 | 1 | 5 | 6 | 0 | 6 | 2225.66230 | 2225.66259 | -0.000289 | 1.00 |
| 4 | 3 | 2 | 5 | 3 | 3 | 2225.66735 | 2225.66811 | -0.000761 | 0.25 |
| 4 | 3 | 1 | 5 | 3 | 2 | 2225.66735 | 2225.66704 | 0.000309 | 0.25 |
| 9 | 2 | 7 | 9 | 3 | 6 | 2225.68135 | 2225.68167 | -0.000324 | 1.00 |
| 5 | 1 | 5 | 5 | 2 | 4 | 2225.68894 | 2225.68908 | -0.000135 | 1.00 |
| 4 | 0 | 4 | 5 | 0 | 5 | 2225.69320 | 2225.69307 | 0.000137 | 1.00 |
| 4 | 1 | 4 | 5 | 1 | 5 | 2225.70667 | 2225.70684 | -0.000166 | 1.00 |
| 3 | 0 | 3 | 4 | 1 | 4 | 2225.71065 | 2225.71080 | -0.000149 | 1.00 |
| 4 | 1 | 4 | 4 | 2 | 3 | 2225.71914 | 2225.71908 | 0.000062 | 1.00 |
| 3 | 1 | 2 | 4 | 1 | 3 | 2225.73427 | 2225.73429 | -0.000018 | 1.00 |
| 3 | 2 | 2 | 4 | 2 | 3 | 2225.75384 | 2225.75372 | 0.000119 | 1.00 |
| 2 | 0 | 2 | 3 | 1 | 3 | 2225.76973 | 2225.76977 | -0.000043 | 1.00 |
| 3 | 1 | 3 | 4 | 1 | 4 | 2225.77872 | 2225.77882 | -0.000098 | 1.00 |
| 5 | 2 | 4 | 6 | 1 | 5 | 2225.78407 | 2225.78423 | -0.000157 | 1.00 |
| 2 | 1 | 1 | 2 | 2 | 0 | 2225.79366 | 2225.79341 | 0.000249 | 1.00 |
| 3 | 1 | 2 | 3 | 2 | 1 | 2225.80589 | 2225.80583 | 0.000056 | 1.00 |
| 6 | 0 | 6 | 6 | 1 | 5 | 2225.81229 | 2225.81219 | 0.000097 | 1.00 |
| 2 | 1 | 2 | 3 | 1 | 3 | 2225.85140 | 2225.85147 | -0.000066 | 1.00 |
| 5 | 0 | 5 | 5 | 1 | 4 | 2225.86516 | 2225.86523 | -0.000071 | 1.00 |
| 4 | 2 | 3 | 5 | 1 | 4 | 2225.89018 | 2225.89045 | -0.000275 | 1.00 |
| 1 | 0 | 1 | 2 | 0 | 2 | 2225.91523 | 2225.91498 | 0.000253 | 1.00 |
| 2 | 1 | 2 | 3 | 0 | 3 | 2225.92010 | 2225.92011 | -0.000016 | 1.00 |
| 1 | 1 | 1 | 2 | 1 | 2 | 2225.92471 | 2225.92469 | 0.000017 | 1.00 |
| 3 | 0 | 3 | 3 | 1 | 2 | 2225.93630 | 2225.93633 | -0.000032 | 1.00 |
| 2 | 0 | 2 | 2 | 1 | 1 | 2225.95653 | 2225.95654 | -0.000016 | 1.00 |
| 3 | 2 | 1 | 3 | 2 | 2 | 2226.07532 | 2226.07522 | 0.000098 | 1.00 |
| 2 | 1 | 1 | 2 | 1 | 2 | 2226.10629 | 2226.10596 | 0.000330 | 1.00 |
| 1 | 0 | 1 | 0 | 0 | 0 | 2226.15194 | 2226.15206 | -0.000124 | 1.00 |
| 1 | 1 | 0 | 1 | 0 | 1 | 2226.17603 | 2226.17604 | -0.000003 | 1.00 |
| 2 | 1 | 1 | 2 | 0 | 2 | 2226.18823 | 2226.18826 | -0.000027 | 1.00 |
| 3 | 1 | 2 | 3 | 0 | 3 | 2226.20835 | 2226.20832 | 0.000028 | 1.00 |
| 4 | 1 | 3 | 4 | 0 | 4 | 2226.23804 | 2226.23797 | 0.000074 | 1.00 |
| 5 | 1 | 4 | 5 | 0 | 5 | 2226.27812 | 2226.27875 | -0.000629 | 1.00 |
| 3 | 1 | 3 | 2 | 1 | 2 | 2226.29250 | 2226.29247 | 0.000031 | 1.00 |
| 3 | 0 | 3 | 2 | 0 | 2 | 2226.30724 | 2226.30676 | 0.000478 | 1.00 |
| 6 | 1 | 5 | 6 | 0 | 6 | 2226.33153 | 2226.33120 | 0.000325 | 1.00 |
| 3 | 2 | 1 | 3 | 1 | 2 | 2226.33670 | 2226.33672 | -0.000021 | 1.00 |
| 8 | 2 | 6 | 8 | 1 | 7 | 2226.34560 | 2226.34523 | 0.000370 | 1.00 |
| 6 | 1 | 5 | 5 | 2 | 4 | 2226.35736 | 2226.35769 | -0.000338 | 1.00 |
| 4 | 1 | 4 | 3 | 1 | 3 | 2226.36475 | 2226.36479 | -0.000032 | 1.00 |
| 4 | 1 | 3 | 3 | 1 | 2 | 2226.40981 | 2226.40991 | -0.000098 | 1.00 |
| 8 | 3 | 5 | 8 | 2 | 6 | 2226.47608 | 2226.47642 | -0.000340 | 1.00 |
| 7 | 3 | 4 | 7 | 2 | 5 | 2226.49869 | 2226.49871 | -0.000018 | 1.00 |
| 6 | 1 | 6 | 5 | 1 | 5 | 2226.50694 | 2226.50701 | -0.000065 | 1.00 |
| 2 | 2 | 0 | 1 | 1 | 1 | 2226.53036 | 2226.53044 | -0.000072 | 1.00 |
| 5 | 3 | 2 | 5 | 2 | 3 | 2226.53684 | 2226.53658 | 0.000257 | 1.00 |
| 6 | 2 | 5 | 5 | 2 | 4 | 2226.54262 | 2226.54168 | 0.000944 | 1.00 |
| 3 | 3 | 0 | 3 | 2 | 1 | 2226.55491 | 2226.55460 | 0.000309 | 1.00 |
| 5 | 3 | 3 | 5 | 2 | 4 | 2226.56673 | 2226.56665 | 0.000071 | 1.00 |
| 7 | 1 | 7 | 6 | 1 | 6 | 2226.57657 | 2226.57692 | -0.000354 | 1.00 |
| 3 | 2 | 2 | 2 | 1 | 1 | 2226.58646 | 2226.58620 | 0.000258 | 0.25 |
| 7 | 0 | 7 | 6 | 0 | 6 | 2226.58646 | 2226.58658 | -0.000121 | 0.25 |
| 7 | 1 | 7 | 6 | 0 | 6 | 2226.60362 | 2226.60396 | -0.000343 | 1.00 |
| 7 | 2 | 6 | 6 | 2 | 5 | 2226.61777 | 2226.61767 | 0.000105 | 1.00 |
| 3 | 2 | 1 | 2 | 1 | 2 | 2226.62477 | 2226.62484 | -0.000070 | 1.00 |
| 8 | 0 | 8 | 7 | 1 | 7 | 2226.63516 | 2226.63520 | -0.000044 | 1.00 |
| 7 | 2 | 5 | 6 | 2 | 4 | 2226.65815 | 2226.65829 | -0.000134 | 1.00 |
| 8 | 1 | 8 | 7 | 0 | 7 | 2226.66395 | 2226.66390 | 0.000049 | 1.00 |
| 9 | 1 | 8 | 8 | 2 | 7 | 2226.67326 | 2226.67344 | -0.000181 | 1.00 |
| 8 | 2 | 7 | 7 | 2 | 6 | 2226.69252 | 2226.69257 | -0.000043 | 1.00 |
| 9 | 0 | 9 | 8 | 1 | 8 | 2226.70805 | 2226.70811 | -0.000053 | 1.00 |
| 9 | 1 | 9 | 8 | 1 | 8 | 2226.71462 | 2226.71479 | -0.000171 | 1.00 |
| 7 | 4 | 3 | 7 | 3 | 4 | 2226.73713 | 2226.73761 | -0.000483 | 1.00 |
| 6 | 2 | 5 | 5 | 1 | 4 | 2226.75643 | 2226.75648 | -0.000049 | 1.00 |
| 9 | 2 | 8 | 8 | 2 | 7 | 2226.76645 | 2226.76631 | 0.000138 | 1.00 |
| 10 | 0 | 10 | 9 | 1 | 9 | 2226.77921 | 2226.77900 | 0.000213 | 1.00 |
| 3 | 3 | 0 | 2 | 2 | 1 | 2226.79695 | 2226.79744 | -0.000485 | 0.25 |
| 3 | 3 | 1 | 2 | 2 | 0 | 2226.79695 | 2226.79642 | 0.000534 | 0.25 |
| 7 | 2 | 6 | 6 | 1 | 5 | 2226.80360 | 2226.80356 | 0.000040 | 1.00 |
| 4 | 3 | 2 | 3 | 2 | 1 | 2226.87373 | 2226.87365 | 0.000080 | 1.00 |
| 4 | 3 | 1 | 3 | 2 | 2 | 2226.87874 | 2226.87881 | -0.000070 | 1.00 |
| 5 | 3 | 2 | 4 | 2 | 3 | 2226.96279 | 2226.96273 | 0.000060 | 1.00 |
| 4 | 4 | 0 | 3 | 3 | 1 | 2227.06853 | 2227.06896 | -0.000426 | 0.25 |
| 4 | 4 | 1 | 3 | 3 | 0 | 2227.06853 | 2227.06890 | -0.000371 | 0.25 |
| 7 | 3 | 5 | 6 | 2 | 4 | 2227.07595 | 2227.07580 | 0.000152 | 1.00 |
| 5 | 4 | 2 | 4 | 3 | 1 | 2227.14806 | 2227.14820 | -0.000140 | 0.25 |
| 5 | 4 | 1 | 4 | 3 | 2 | 2227.14806 | 2227.14858 | -0.000525 | 0.25 |
| 6 | 4 | 2 | 5 | 3 | 3 | 2227.22764 | 2227.22836 | -0.000718 | 0.25 |
| 6 | 4 | 3 | 5 | 3 | 2 | 2227.22764 | 2227.22681 | 0.000830 | 0.25 |
| 7 | 4 | 3 | 6 | 3 | 4 | 2227.30857 | 2227.30860 | -0.000030 | 1.00 |
| 5 | 5 | 1 | 4 | 4 | 0 | 2227.34004 | 2227.34009 | -0.000058 | 1.00 |
| 5 | 5 | 0 | 4 | 4 | 1 | 2227.34004 | 2227.34010 | -0.000061 | 1.00 |
| 8 | 4 | 5 | 7 | 3 | 4 | 2227.37881 | 2227.37840 | 0.000410 | 1.00 |
| 8 | 4 | 4 | 7 | 3 | 5 | 2227.38984 | 2227.38995 | -0.000111 | 1.00 |
| 6 | 5 | 2 | 5 | 4 | 1 | 2227.41948 | 2227.41958 | -0.000104 | 0.25 |
| 6 | 5 | 1 | 5 | 4 | 2 | 2227.41948 | 2227.41960 | -0.000126 | 0.25 |
| 9 | 4 | 6 | 8 | 3 | 5 | 2227.44869 | 2227.44848 | 0.000209 | 1.00 |
| 7 | 5 | 2 | 6 | 4 | 3 | 2227.49894 | 2227.49900 | -0.000060 | 0.25 |
| 7 | 5 | 3 | 6 | 4 | 2 | 2227.49894 | 2227.49889 | 0.000050 | 0.25 |
| 8 | 5 | 3 | 7 | 4 | 4 | 2227.57826 | 2227.57826 | -0.000004 | 0.25 |
| 8 | 5 | 4 | 7 | 4 | 3 | 2227.57826 | 2227.57786 | 0.000401 | 0.25 |

*****************************************************************